\documentclass[aps,pra,reprint,twocolumn,amsmath,amssymb,showpacs]{revtex4-1}
\usepackage{graphicx}
\usepackage[bookmarks,bookmarksopen,bookmarksnumbered,colorlinks,linkcolor=red,linktocpage,citecolor=blue,urlcolor=cyan,pdfpagemode=UseOutline]{hyperref}
\usepackage{epstopdf}

\def\calE{{\cal E}}

\def\brp{\bar\br}

\def\n{n}
\def\br{{\bf r}}
\def\ben{\begin{equation}}
\def\een{\end{equation}}

\newcommand{\intd}{\mathrm{d}}
\newcommand{\parref}[1]{(\ref{#1})}
\newcommand{\vect}[1]{\mathbf{#1}}
\newcommand{\matelem}[3]{\left\langle #1 \left| #2 \right| #3 \right\rangle}
\providecommand{\abs}[1]{\left|#1\right|}
\DeclareMathOperator{\erf}{erf}
\DeclareMathOperator{\erfc}{erfc}
\DeclareMathOperator{\sgn}{sgn}

\begin{document}
\title{Non-existence of Taylor expansion in time due to cusps}
\author{Zeng-hui Yang}
\affiliation{Department of Physics and Astronomy, University of Missouri - Columbia, Columbia, MO 65211, USA}
\author{Kieron Burke}
\affiliation{Department of Chemistry, University of California - Irvine, Irvine, CA 92697, USA}
\date{\today}

\begin{abstract}
In the usual treatment of electronic structure, all matter
has cusps in the electronic density at nuclei. Cusps can 
produce non-analytic behavior in time, even in response to perturbations that
are time-analytic. We analyze these non-analyticities in a
simple case from many perspectives. We describe a method,
the $s$-expansion, that can be used in several such cases,
and illustrate it with a variety of examples. These include both the 
sudden appearance of electric fields and disappearance of nuclei, in both one
and three dimensions. 
When successful, the $s$-expansion yields the dominant short-time behavior, no matter
how strong the external electric field, but agrees with linear response
theory in the weak limit.
We discuss the relevance of these
results to time-dependent density functional theory.
\end{abstract}

\maketitle

\section{Introduction}
\label{sect:intro}
Time-dependent quantum mechanics is used to calculate the response
of systems to time-varying external potentials\cite{T07},
but can be computationally demanding for many particles.  Among
practical methods,
time-dependent density functional theory (TDDFT) excels as a computationally
inexpensive method for dealing with the interactions between electrons
in time-dependent quantum mechanics\cite{MMNG12,U12}.
In the last two decades,
use of TDDFT has grown tremendously, especially
for calculations of transition frequencies of electronic excitations
in molecules\cite{EBF09} and solids\cite{ORR02,BSDR07}.
The relative computational ease
with which TDDFT handles electron-electron interaction make it the
only viable quantum tool for systems with several hundred atoms\cite{MMNG12,EBF09,U12}.

But the validity of TDDFT relies on 
the celebrated Runge-Gross theorem\cite{RG84} which proves, under certain
circumstances, that the time-dependent one-body potential of
an interacting electronic system is a functional of the one-electron
density.   Modern TDDFT calculations also use a Kohn-Sham scheme,
in which fictitious non-interacting fermions are propagated in
a time-dependent multiplicative potential, defined to reproduce the time-dependent
density of the interacting system.   
Use of such a scheme implicitly supposes that such a potential
exists (in technical jargon, that the density is non-interacting
$v$-representable\cite{MMNG12,U12}). Ground-breaking work by
van Leeuwen\cite{L99} showed that, under quite general
assumptions, such a potential can always be found, apparently
ending this question within TDDFT.

However, nature can occasionally be both subtle and malicious.
The Runge-Gross theorem assumes time-Taylor expandability($t$-TE) of the
time-dependent potential, while van Leeuwen's proof requires 
such expandability of the density also.  In recent work\cite{YMB12}, we gave
a very simple, realistic case (a hydrogen atom in a suddenly-switched
static electric field) in which the latter fails,
thus reopening the issue of $v$-representability in TDDFT.
This could only be done convincingly by creating a methodology
for explicitly extracting the short-time asymptotic behavior in
such cases, and demonstrating the non-expandability of the density.
This has reopened the question of the existence of a KS potential
in the common case of Coulomb attraction to the nuclei, and
recent work has focused on avoiding the Taylor expansion in time\cite{RL11,RGPL12}.

These results were quite unexpected, as they are due to the
non-interchangeability of two commonly interchanged limits.
In fact, as we demonstrate explicitly here, the time-dependent
density in such cases, $\n(\br,t)$, has {\em no} well-defined
short time expansion.  For finite distances from a cusp, one
asymptotic expansion applies, while for distances less than ${\sqrt{t}}$
from a cusp, a different expansion dominates.
(Atomic units $e=\hbar=m_e=1/(4\pi\epsilon_0)=1$ are used throughout.)
A related statement
is that we find the radius of expansion of the time-Taylor series is 
0.  However, even if the density has no well-defined expansion,
integrals over the density, such as the time-dependent dipole moment,
{\em are} well-defined, but can contain fractional powers of $t$.
Here we give further examples of the method developed in Ref. \cite{YMB12} for
calculating some of these quantities for several cases.
We also show how these features appear in various alternative
approaches to this problem.

Our work here is far from a complete analysis of these behaviors,
and
we make no attempt at a general treatment of this problem.  Instead,
we merely scratch the surface of the very thorny issues created
by the coupling between space and time in the Schr\"odinger
equation.  We hope this work will inspire more comprehensive study
of these questions, and perhaps lead to a more straightforward
computational scheme.

The paper is divided as follows.  We begin by analyzing a very simple
illustration, the 1D disappearing nucleus, from many different viewpoints.  
Although this is not a three-dimensional
Coulomb potential problem, this illustration is chosen
because we have closed analytic results. 
We next present the $s$-expansion as a general method for extracting
the short-time behavior of these systems.  We then revisit 1D.
We check our method reproduces the analytic results of the disappearing
nucleus problem, and show what it produces for a nucleus in an electric field.
We then turn to 3D, applying the method to the two previous cases, but in 3D.
There are specific complications for the H atom in an electric field.
In the following section, we examine the time-dependent dipole moment,
rather than just the wavefunction, finding its behaviour entirely in
the disappearing nucleus case, and partially in the electric-field problem.
Then we discuss more general potentials in space and time
(but not any general class of potentials).
We close with a discussion of the implications of these results for
many-electron systems and time-dependent density functional theory.

\section{When nuclei vanish}
\label{sect:illus}
Here we study the failure of the Taylor series in the simplest possible
case, first studied in Ref. \cite{MTWB10}.  In one dimension, we begin at $t=0$ with a wavefunction:
\begin{equation}
\psi_0(x)=\exp(-\abs{x}),
\label{eqn:intro:psiinit}
\end{equation}
which has a cusp at $x=0$.
We propagate with the free-particle Hamiltonian
\begin{equation}
\hat{H}=-\frac{1}{2}\frac{d^2}{dx^2},
\label{eqn:intro:H}
\end{equation}
and find 
\ben
\psi(x,t>0)=\hat{U}(t)\, \psi_0(x),
\label{eqn:intro:psit}
\een
where the time-propagation operator $\hat{U}(t)=\exp(-i\hat{H}t)$ because the Hamiltonian is $t$-independent.
The common trick of $t$-TE uses:
\ben
\exp[-i\hat{H}t]\stackrel{?}{=}1-i\hat{H}t-\hat{H}^2t^2/2+\cdots.
\label{eqn:intro:tTE}
\een
Many textbooks either use the $t$-TE interchangeably with the correct
spectral definition of the propagator\cite{AGD75,FW03,M95,BF04}, or
introduce the $t$-TE as a formal propagation method without
further discussion of the implications\cite{T07}. 
For the 1D example system, we evaluate the time-dependent wavefunction with
the Taylor-expanded time-evolution operator:
\ben
\begin{split}
\psi^\text{TE}(x,t>0)&
=\left[\sum_{j=0}^\infty(-i\hat{H})^jt^j/j! \right]\,\psi_0(x)\\
&=\exp(-\abs{x}+i t/2),\quad\quad\quad\quad(x\ne0).
\label{eqn:1dnucdisapp:TEwf}
\end{split}
\een
yielding the remarkable result that (for $x\neq 0$), the density appears to remain stationary!

We refer to this example as a 1d vanishing nucleus, because the
initial wavefunction 
is the eigenstate of $v(x)=-\delta (x)$, and decays
exponentially like that of a hydrogen atom.
According to the Taylor expansion,
we can instantly remove this potential at $t=0$, and the density does not
change.  Obviously, if we do nothing to the potential, the density will not
change either, in apparent contradiction of the RG theorem.

\begin{figure}[htbp]
\includegraphics[height=\columnwidth,angle=-90]{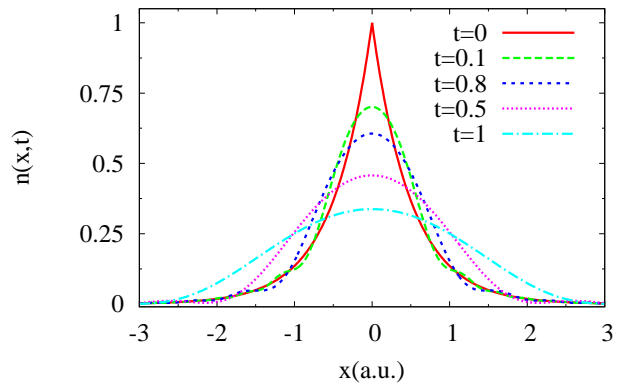}
\caption{(Color online) Time-dependent density of an exponential that propagates freely after
$t=0$.  The Taylor expansion fails to change from its $t=0$ value.}
\label{fig:intro:nucdisapp}
\end{figure}

In this case, it is simple to find the true wavefunction.  
The free-particle propagator in 1d is
\ben
\begin{split}
U(x,x',t>0)&=\frac{i}{2\pi}\int_{-\infty}^\infty\intd k\;\exp[iku-ik^2t/2]\\
&=\frac{\exp[iu^2/(2t)]}{\sqrt{2\pi i t}},
\end{split}
\label{eqn:intro:timeevoloper}
\een
where $u=x-x'$,
and convolution with $\psi_0(x)$ yields
\ben
\psi(x,t>0)={\cal S}_x\left[\exp(x+i t/2)\erfc\left(\frac{x+it}{\sqrt{2it}}\right)\right]
\label{eqn:1dnucdisapp:exactwf}
\een
and ${\cal S}_x\, f = [f(x)+f(-x)]/2$ extracts the spatially symmetric part
of a function. We choose $\sqrt{i}=(1+i)/\sqrt{2}$ and use this branch through the paper for square roots.
Fig. \ref{fig:intro:nucdisapp} confirms that the wavefunction spreads
and the cusp vanishes for $t>0$, as intuition demands. 
An important feature of Eq. \parref{eqn:1dnucdisapp:exactwf} is that $\psi(x,t)$ is not
an analytic function at $t=0$ with respect to $t$, and 
we denote this as time-non-analyticity throughout the paper. 
We analyze the time-non-analyticity in detail in Sect. \ref{sect:method}.

In practice, we observe that the $t$-TE wavefunction works when the initial wavefunction is space-analytic, from which wavefunctions with cusps are excluded. We define `cusp' in a general sense as a discontinuity in the space derivatives of a certain order of the concerned function. No matter what the external potential is, a Hamiltonian always contains the kinetic energy operator - a differential operator in space. According to Eq. \parref{eqn:1dnucdisapp:TEwf},
the existence of a $t$-TE wavefunction requires the initial wavefunction being differentiable to infinite order at any space point. However, infinite-order differentiable does not guarantee the validity of the $t$-TE wavefunction: the 
wavefunction is differentiable in the distributional sense
in the case of cusps; in another case\cite{HS72},
one can construct a non-trivial wavepacket where the space derivative
of all orders vanish at certain points; $t$-TE fails in both cases.
The analyticity of the initial wavefunction in space is linked to the analyticity of the TD wavefunction in time. We provide more evidence and a heuristic derivation of time-non-analyticities originating from cusps in Sect. \ref{sect:app:tnonanalyfromcusp}.

For simplicity of notations, all time variables are greater than 0 unless otherwise specified.

\subsection{Interchanging orders of limits}
The failure of $t$-TE is due to the interchange of the order
of limiting operations.  For a time-independent Hamiltonian\cite{YMB12},
\ben
\psi(\br,t)
=\sum_j c_j\left(\sum_{p=0}^\infty\frac{(-i\epsilon_j)^p}{p!}t^p\right)\phi_j(\br),
\label{eqn:TE:exactwf}
\een
in which $c_j=\langle \phi_j | \psi_0 \rangle$, while
\ben
\psi^\text{TE}(\br,t)=\sum_{p=0}^\infty \left(\sum_j c_j \frac{(-i\epsilon_j)^p}{p!}\phi_j(\br)\right)t^p,
\label{eqn:TE:TEwf}
\een
which is obtained by interchanging the order of the two summations. 
If the initial wavefunction is composed
of a finite number of eigenstates, 
such an interchange is valid.  More generally,
one requires uniform convergence for two summations of infinite number of terms to be interchangeable.

We now perform a $t$-TE on the integrand of Eq. \parref{eqn:intro:timeevoloper}
and interchange the order of the integration and the summing of $t$-TE:
\ben
\begin{split}
U^\text{TE}(x,x',t)&=\frac{i}{2\pi}\sum_{n=0}^\infty\int_{-\infty}^\infty\intd k\;\exp[iku]\frac{(-ik^2/2)^n}{n!}t^n\\
&=\sum_{n=0}^\infty\frac{i^n}{2^n\,n!}\delta^{(2n)}(u)t^n,
\end{split}
\een
where $u=x-x'$, and $\delta^{(2n)}$ denotes the $2n$th order derivative of
the $\delta$-function with respect to $u$. Thus $U^\text{TE}$ only exists in a
distributional sense. Applying $\hat{U}^\text{TE}(t)$
to Eq. \parref{eqn:intro:psiinit} generates an ill-defined wavefunction, even in the distributional sense:
\ben
\begin{split}
&\psi^\text{TE}(x,t>0)=\psi_0(x)-it\left[-\frac{1}{2}+\delta(x)\right]\psi_0(x)\\
&\quad-\frac{t^2}{2}\Bigg\{\left[\frac{1}{4}-\frac{1}{2}\delta''(x)-\delta(x)+\delta^2(x)\right]\psi_0(x)\\
&\quad-\delta'(x)\psi'_0(x)\Bigg\}+O(t^3),
\label{eqn:1dnucdisapp:fullTEwf}
\end{split}
\een

$t$-TE does not apply to systems with cusps due to the problematic interchange of limiting operations. In many cases, one can recover the correct result by introducing another interchange of limiting operations. Here, we notice that the initial wavefunction does not have a cusp in momentum space:
\ben
\begin{split}
\Psi_0(k)&=\int_{-\infty}^\infty\intd x\;\psi_0(x)\exp[-ikx]\\
&=\frac{2}{k^2+1},
\end{split}
\label{eqn:intro:psi0momentum}
\een
where we denote the Fourier transform of $\psi$ with respect of $x$ as $\Psi$ and the conjugate variable of $x$ as $k$.

According to our previous argument, $t$-TE should be valid for this case. By applying $U^\text{TE}$ to Eq. \parref{eqn:intro:psi0momentum} and performing the summation to infinite order of the $t$-TE, we obtain the $t$-TE wavefunction in momentum space as
\ben
\begin{split}
\psi^\text{TE}(k,t)&=\sum_{n=0}^\infty\frac{(-ik^2/2)^n}{n!}\psi_0(k)t^n\\
&=\frac{2}{k^2+1}\exp(-ik^2t/2),
\end{split}
\een
which is exactly the Fourier transform of Eq. \parref{eqn:1dnucdisapp:exactwf}, the correct TD wavefunction. Taking the $t$-TE in momentum space is equivalent to performing a Fourier transform on Eq. \parref{eqn:1dnucdisapp:fullTEwf}, and then interchanging the order of the Fourier transform with the summation of the $t$-TE series. By introducing this extra interchange of orders, the correct TD behaviors are recovered. The Borel summation of asymptotic series(as in Sect. \ref{sect:app:1ddisappnuc}) is another example of correcting the wrong result from interchanging the order of limiting operations by introducing another interchange of orders, and we develop in Sect. \ref{sect:method} a method based on the Borel summation to obtain short-time behaviors for systems with cusps. Unfortunately, there is no general theorem about the applicability of such techniques, and this topic remains under active research.\cite{BO99,W10}

\subsection{Inner and outer regions}
\label{sect:innerandouter}

\begin{figure}[htbp]
\centering
\includegraphics[height=\columnwidth,angle=-90]{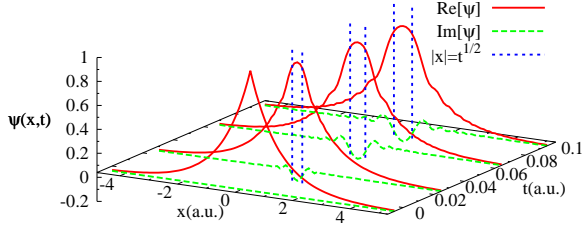}
\caption{(Color online) The time-dependent wavefunction after the nucleus vanishes.}
\label{fig:1dnucdisapp:analysisxt}
\end{figure}

Here we define carefully the inner and outer regions, each of which
has a distinct asymptotic expansion.
Fig. \ref{fig:1dnucdisapp:analysisxt} shows the region far from the
origin becomes oscillatory, showing the plane-wave nature
of the eigenstates of the free-particle Hamiltonian; 
yet the region near the origin is non-oscillatory, resembling the spread-out cusp.
By carefully taking the $t\to0_+$ limit as below, 
we notice $t\to0_+$ actually corresponds to two different limits, 
with $\abs{x}\gg\sqrt{t}$ and $\abs{x}\ll\sqrt{t}$ respectively. 
We denote the $\abs{x}\gg\sqrt{t}$ region as the outer region, and 
the $\abs{x}\ll\sqrt{t}$ region as the inner region. 
The correct short-time behavior is composed of the short-time behaviors of these two regions.

\begin{figure}[htbp]
\centering
\includegraphics[height=\columnwidth,angle=-90]{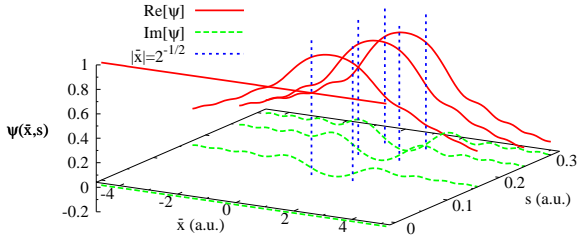}
\caption{(Color online) Same as Fig 2, but as a function of $s=\sqrt{t}$ and $\bar x
=x/\sqrt{2t}$.}
\label{fig:1dnucdisapp:analysissy}
\end{figure}

The short-time series expansions for these two regions can be obtained by changing the variables from $(x,t)$ to the 
following reduced variables:
\ben
s=\sqrt{t},\quad \bar x=\frac{x}{\sqrt{2t}}.
\een
Fig. \ref{fig:1dnucdisapp:analysissy} shows such a 
change-of-variables effectively zooms in to the inner region, and the cusp
in the initial wavefunction is removed in the reduced variables.

We can analytically extract the functions describing the smooth and oscillatory parts of the wavefunction.  Define the following special functions:
\ben
\begin{split}
\mathrm{E_c}(s,{\bar x})&={\cal S}_{\bar x}\left[e^{\sqrt{2}s{\bar x}}\erfc\left(\frac{s_+}{2}\right)\right],\\
\mathrm{F}(s,{\bar x})&={\cal S}_{\bar x}\left[\frac{2}{\sqrt{\pi}}e^{-s_-^2/4}\int_0^{s_-/2}\intd t\; e^{t^2+is_+t}\right],
\end{split}
\een
where $s_\pm=s\pm\sqrt{2}{\bar x}$.

The TD wavefunction Eq. \parref{eqn:1dnucdisapp:exactwf} is then
\ben
\psi(x,t)=\frac{1}{2}e^{\frac{is^2}{2}}\left[\mathrm{E_c}\left(s,{\bar x}\right)-i\mathrm{F}^*(s,{\bar x})\right].
\label{eqn:theory:psispecial}
\een
For the wavefunction, the $\mathrm{E_c}$ part is smooth, and the $-i\mathrm{F}^*$ part oscillates. Since at $t=0$ the oscillatory part does not exist, it must be the effect of the vanishing cusp. In terms of error functions\cite{AS72}:
\ben
\mathrm{F}(s,\bar x)=i e^{\sqrt{2}s{\bar x}}\left\{\erf\left[\sqrt{i}\left({\bar x}-\frac{is}{\sqrt{2}}\right)\right]-\erf\left(\frac{s+\sqrt{2}{\bar x}}{2}\right)\right\}.
\label{eqn:illus:regions:funf}
\een
If $\abs{x}\gg\sqrt{t}$ as $t\to0_+$, the arguments of
the error functions in Eq. \parref{eqn:illus:regions:funf} approach $\infty$. 
On the other hand, when $\abs{x}\ll\sqrt{t}$ as $t\to0_+$, these arguments approach $0$.

The inner-region expansion can be obtained by Taylor-expanding $\psi(s,\bar x)$ as $s\to0_+$ while holding $\bar x$ fixed:
\ben
\begin{split}
&\psi^\text{inner}(s,\bar{x})\stackrel{s\to0_+}{\sim}1+s\left[-\sqrt{2i/\pi}\exp(i \bar{x}^2)\right.\\
&\left.\quad\quad\quad\quad\quad\quad\quad-\sqrt{2}\bar{x}\erf\left(\sqrt{i}^*\bar{x}\right)\right]+O(s^2)\\
&=1+\frac{x^2}{2}-x\erf\left(\sqrt{\frac{i}{2t}}^*x\right)-\sqrt{\frac{2i}{\pi}}\exp\left(\frac{ix^2}{2t}\right)\sqrt{t}+\cdots.
\label{eqn:1dnucdisapp:inner}
\end{split}
\een
The outer-region expansion can be obtained by expanding $\psi(s,\bar x)$ as $\bar x\to\pm\infty$ while holding $s$ fixed:
\ben
\begin{split}
&\psi^\text{outer}(s,\bar{x})\stackrel{\bar x\to\pm\infty}{\sim}\exp(-\sqrt{2}s\abs{\bar x}+is^2/2)\\
&\quad\quad\quad+\sqrt{\frac{i}{2\pi}}^*\exp(i{\bar x}^2)s{\bar x}^{-2}+O({\bar x}^{-4})\\
&=\exp(-\abs{x})\left(1+\frac{it}{2}\right)+\sqrt{\frac{2i}{\pi}}^*\frac{\exp[ix^2/(2t)]}{x^2}t^{3/2}+\cdots
\label{eqn:1dnucdisapp:outer}
\end{split}
\een

\begin{figure}[htbp]
\centering
\includegraphics[height=\columnwidth,angle=-90]{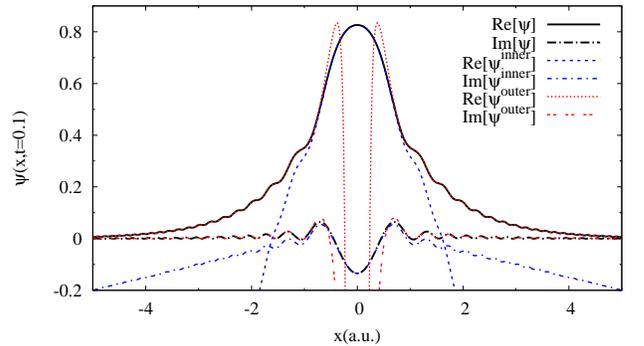}
\caption{(Color online) $\psi(x,t=0.1)$ after the nucleus vanishes, 
plotting the truncated inner-region expansion and the outer-region expansion.}
\label{fig:1dnucdisapp:innerandouter}
\end{figure}

The truncated inner-region and outer-region expansions are plotted in Fig. \ref{fig:1dnucdisapp:innerandouter}. These two together define the correct short-time behavior of the wavefunction. It should be noted that the usual $t\to0_+$ limit only corresponds to the outer region, and thus it does not contain all the information of the system at $t\to0_+$. Since these two expansions both contain time-non-analyticities, such as half-powers in $t$ and $\exp[ix^2/(2t)]$, the $t$-TE cannot describe the correct short-time behavior.

Note that the outer region expansion can also be found via the stationary phase approximation\cite{BO99},
applied to the propagated wavefunction in momentum space. The stationary phase approximation is a method yielding the leading asymptotic behavior (as $\xi\to\infty$) of integrals of the following form:
\ben
\label{eqn:101229:statphaseint}
I(\xi)=\int_a^b\intd \zeta\;f(\zeta)\exp[i\xi g(\zeta)].
\een
Write out the time-dependent wavefunction using the Green's function\cite{E06}:
\ben
\begin{split}
\psi(x,t)&=\frac{i}{2\pi}\int\;k\intd k\int\intd x'\:\exp(-i k^2t/2)\tilde{G}(x,x',\frac{k^2}{2})\psi_0(x')\\
&=\frac{1}{\pi}\int_0^\infty\intd k\;\left[\frac{i k\exp(-\abs{x})}{k^2+1}+\frac{\exp(i k\abs{x})}{k^2+1}\right]\\
&\quad\quad\quad\times\exp(-i k^2t/2),
\label{eqn:statphase:1dnucdisapp}
\end{split}
\een
where the time-domain Green's function $G$ is related to the time-propagation operator introduced in Sect. \ref{sect:illus} by $G(x,x',t)=-i\matelem{x}{\hat{U}(t)}{x'}$, and $\tilde{G}$ is its Fourier transform with respect to $t$. Eq. \parref{eqn:statphase:1dnucdisapp} is equivalent to Eq. \parref{eqn:intro:psit}.

The first term in the integral gives the $t$-TE wavefunction, and we
apply the stationary phase approximation\cite{BO99}.
The stationary point is $k=\abs{x}/t$, and the second term of Eq. \parref{eqn:statphase:1dnucdisapp} correctly yields the leading time-non-analyticity:
\ben
\psi(x,t)\stackrel{x/\sqrt{2t}\to\infty}{\sim} \cdots+\sqrt{\frac{2i}{\pi}}^*\frac{\exp[i x^2/(2t)]}{x^2}t^{3/2}+\cdots.
\label{eqn:statphase:1douter}
\een

By change-of-variables, one can see the limit that the stationary phase approximation corresponds to. Letting $k=(\abs{x}/t)\zeta$, the second term in Eq. \parref{eqn:statphase:1dnucdisapp} becomes
\ben
\int_0^\infty\intd \zeta\;\exp\left[i\frac{x^2}{2t}(-\zeta^2+2\zeta)\right]\frac{\abs{x}t}{x^2\zeta^2+t^2}.
\label{eqn:statphase:transformed}
\een
Eq. \parref{eqn:statphase:transformed} is in the form of Eq. \parref{eqn:101229:statphaseint}, with $x^2/(2t)$ as $\xi$ in Eq. \parref{eqn:101229:statphaseint}. Thus the stationary phase approximation Eq. \parref{eqn:statphase:1douter} corresponds to the $x^2/(2t)\to\infty$ limit, i.e., the outer-region expansion.

\subsection{Radius of convergence}
\label{sect:illus:radconv:1ddisappnuc}
Next we consider the radius of convergence of the Taylor expansion. 
We study the wavefunction in the vanishing nucleus problem, beginning at $t_0$ {\em after}
the nucleus vanishes.  This wavefunction has no cusp, and has a well-behaved
Taylor expansion.

The $t$-TE of the time-evolution operator is
\ben
\mathrm{TE}\left[e^{-iHt}\right]=\sum_j\frac{(it/2)^j}{j!}\frac{\partial^{2j}}{\partial x^{2j}}.
\een
The exact TD wavefunction is Eq. \parref{eqn:1dnucdisapp:exactwf}. Instead of $t$-TE at $t=0$, we pick a later time $t_0$ as the expansion point, and derive the radius of convergence of this $t$-TE.

In the outer region($x\gg\sqrt{t_0}$), it is easy to show that
\ben
\frac{\partial^{2n}\psi(x,t)}{\partial x^{2n}}\propto-\frac{x^{2n-2}}{t_0^{2n-3/2}}C,\quad\quad x\gg\sqrt{t_0},
\een
where $C=(i-1)e^{ix^2/(2t_0)}/\sqrt{\pi}$. In the inner region($x\ll\sqrt{t_0}$), we have
\ben
\frac{\partial^{2n}\psi(x,t)}{\partial x^{2n}}\propto\frac{i^{n-1}(2n-1)!!}{t_0^{n-1/2}}C,\quad\quad x\ll\sqrt{t_0}.
\een
Then the radii of convergence for the inner and outer regions are given by
\ben
R_\text{outer}=\infty,\quad\quad R_\text{inner}=t_0,
\een
separately. Thus the radius of convergence for the inner region vanishes as $t_0\to0$.

\section{The $s$-expansion}
\label{sect:method}
Here we introduce the $s$-expansion method\cite{YMB12}. Our notation is for 3D problems, but the method applies equally to 1D problems.
Based on the previous analysis, we begin by a change-of-variables:
\ben
s=\sqrt{t},\quad \brp=\frac{\br}{\sqrt{2t}}.
\label{eqn:method:reducedvar}
\een
With these reduced variables, we can describe the time-non-analyticities which are not covered by the form of the $t$-TE. The time-dependent Schr\"odinger equation becomes
\ben
\bar\nabla^2 \psi-4s^2 v\,\psi+2i\left\{s\frac{\partial\psi}{\partial s}-\brp\cdot\bar\nabla\psi\right\}=0.
\label{eqn:method:TDSEred}
\een
In the vanishing nucleus case, $\psi$ is equal to its Taylor expansion in powers of $s$ for fixed $\brp$, and thus we assume the following $s$-expansion ansatz in the more general case:
\ben
\psi(\brp,s) = \sum_{n=0}^\infty\, \psi_{(n)} (\brp)\, s^n.
\label{eqn:method:ansatz}
\een
This yields a set of differential equations:
\ben
\label{eqn:method:diffeqn}
\bar\nabla^2\psi_{(n)}-2i\brp\cdot\bar\nabla\psi_{(n)}+2ni\psi_{(n)}-4\sum_{p=-2}^{n-2} v_{(p)}\psi_{(n-p-2)}=0,
\een
in which we assume the potential has a simple form of $v(\br,t)=\sum_{p=-2}^\infty v_{(p)}(\brp)\, s^p$.  Thus each power of $s$ produces a second-order differential equation for a function of $\brp$. Eq. \parref{eqn:method:diffeqn} requires proper boundary conditions for the solution to be well-defined. Eq. \parref{eqn:method:diffeqn} is equivalent to the TDSE whenever the wavefunction ansatz Eq. \parref{eqn:method:ansatz} is applicable. This requires the boundary conditions to be derived from the initial condition of the TDSE, which is the initial wavefunction $\psi_0(\br)\,\equiv\,\psi_0({\sqrt{2}} \brp s)$. For finite argument $\br$, $s\to 0$ implies $\bar r \to \infty$, so the expansion of the initial wavefunction at $s\to0$ determines the large $\brp$ behavior of the $\psi_{(m)}(\brp)$, i.e., provides the boundary conditions of Eq. \parref{eqn:method:diffeqn}.

We first check that for the trivial case where a system stays in an eigenstate, the $s$-expansion reduces to the $t$-TE result. Assume the system stays in an eigenstate $\phi(\br)$ with eigenvalue $E$, Eq. \parref{eqn:method:TDSEred} becomes
\ben
-4s^2E\psi+2i\left\{s\frac{\partial\psi}{\partial s}-\brp\cdot\bar\nabla\psi\right\}=0,
\een
and Eq. \parref{eqn:method:diffeqn} becomes
\ben
-4E\psi_{(n-2)}+2ni\psi_{(n)}-2i\brp\cdot\bar\nabla\psi_{(n)}=0.
\label{eqn:method:diffeqn:eig}
\een
Eq. \parref{eqn:method:diffeqn:eig} can be trivially solved, and the coefficients originating from the differential equations are determined by the initial condition of the TDSE. Inserting $\psi_{(n)}$'s into Eq. \parref{eqn:method:ansatz}, we obtain
\ben
\psi(\brp,s)=\phi(0)+\sqrt{2}\brp\phi'(0)s+[-iE\phi(0)+\brp^2\phi''(0)]s^2+\cdots,
\label{eqn:method:eigres}
\een
where the derivatives of $\phi$ are taken with respect of $\br$. Eq. \parref{eqn:method:eigres} is identical to the $t$-TE result.

Several examples of using the method are provided in Sect. \ref{sect:app:1ddisappnuc}, \ref{sect:app:1dHinEfield}, and \ref{sect:app:3dHinEfield}. For the 1D vanishing nucleus case, we solve Eq. \parref{eqn:method:diffeqn} directly in Sect. \ref{sect:app:1ddisappnuc}. Partial differential equations as in Eq. \parref{eqn:method:diffeqn} are difficult to solve exactly except for the most simple systems. For the 1D/3D hydrogen in turned-on static electric field shown in Sect. \ref{sect:app:1dHinEfield} and \ref{sect:app:3dHinEfield}, we are not able to solve Eq. \parref{eqn:method:diffeqn} directly. For these more general cases, we find that although $t$-TE does not describe the correct short-time behaviors as a whole, it works fine before the occurrence of the first time-non-analytic term. Thus instead of solving the short-time behaviors directly, we solve for the simpler corrections from the $t$-TE with the method of dominant balance\,(described in Sect. \ref{sect:app:1ddisappnuc}). The correction from the $t$-TE is expressed as asymptotic series. By performing the Borel summation as in Sect. \ref{sect:app:1ddisappnuc}, we obtain the short-time behavior in closed form.

This method is not intended to be applied to all systems. The formulation only applies to one-electron systems. Second, although the theory is applicable for short-time behaviors to any order, the method depends on the ability to solve differential equations analytically in closed form - either directly or through the use of Borel summation - which requires asymptotic expansions in closed form. The requirement of closed-form solutions makes numerical approximation difficult. Although several approximation methods exist for the Borel summation\cite{BO99} requiring only part of the asymptotic series, it is not clear to us whether they are applicable in this case. Third, the short-time behavior obtained from the method is that of the TD wavefunction, which is not an observable. It is usually more important to be able to predict non-analyticities in observables, such as the $\omega^{-7/2}$ in the high frequency oscillator strengths of atoms. However, there is no guarantee that the leading order time-non-analyticity of the TD wavefunction is sufficient to determine that of a desired observable. Sec.  \ref{sect:app:3dHinEfield} and \ref{sect:app:mufromalpha} demonstrate such a situation for the 3D hydrogen atom in a turned-on static electric field. Aside from these restrictions, a more subtle restriction of the method is related to having more than one time-scale introduced by cusps, and is discussed in Sec.  \ref{sect:app:restriction}.

\section{Applications in 1D}
Here we show how the $s$-expansion works, by applying it to several different problems.
We already have the exact solution for a vanishing nucleus, so this works as a demonstration
of our method.

\subsection{Vanishing nucleus revisited}
\label{sect:app:1ddisappnuc}
In this case, Eq. \parref{eqn:method:diffeqn} becomes:
\ben
\label{eqn:101229:diffeqn}
\psi_{(n)}''-2i\bar{x}\psi_{(n)}'+2in\psi_{(n)}=0.
\een
with general solution
\ben
\psi_{(n)}({\bar x})=a_n \mathrm{H}_n(\sqrt{i}{\bar x})+b_n f_n({\bar x}),
\label{eqn:101229:exactsoln}
\een
where
\ben
f_n({\bar x})=\left\{\begin{array}{lr}\mathrm{H}_n(\sqrt{i}{\bar x})\int_0^{\bar x}\intd{\bar x}'\;\frac{\exp[i({\bar x}')^2]}{\mathrm{H}_n(\sqrt{i}{\bar x}')^2} & n\text{ even},\\ _1\mathrm{F}_1\left(-\frac{n}{2},\frac{1}{2},i{\bar x}^2\right) & n\text{ odd,} \end{array}\right.
\een
$\mathrm{H}$ is the Hermite polynomial, and $\mathrm{_1F_1}$ is Kummer's confluent hypergeometric function\cite{AS72}.

Expanding Eq. \parref{eqn:intro:psiinit} at $s\to0_+$ yields
\ben
\psi_0(x)=\exp(-\sqrt{2}s\abs{\bar x})=\sum_{n=0}^\infty\frac{(-\sqrt{2})^n\abs{\bar x}^n}{n!}s^n.
\een
Thus the boundary conditions for Eq. \parref{eqn:101229:exactsoln} are
\ben
\psi_{(n)}(\bar x) \sim (-\sqrt{2})^n |\bar x|^n/n!,~~~|\bar x|\to\infty.
\label{eqn:initcond}
\een
With Eq. \parref{eqn:initcond}, we find
$a_{2n+1}=b_{2n}=0$ and
\ben
\label{eqn:101229:coeff}
a_{2n}=\frac{(-i)^{n}}{(2n)!2^n},~~~~ b_{2n+1} = -\frac{\sqrt{2}i^{n+1/2}}{(2n+1)!!\sqrt{\pi}}.
\een
With Eq. \parref{eqn:101229:exactsoln} and \parref{eqn:101229:coeff}, we obtain the inner-region expansion from Eq. \parref{eqn:method:ansatz}. It agrees with the previously shown Eq. \parref{eqn:1dnucdisapp:inner}, which is obtained from exactly solving the entire TDSE. 

The short-time behavior of the time-dependent wavefunction is described by the inner-region and the outer-region expansion together. The inner-region expansion corresponds to expanding the exact time-dependent wavefunction at $s\to0$ while holding $\brp=\br/\sqrt{2t}$ constant, but there is no requirement on the magnitude of the constant. Therefore the outer-region expansion is obtained by expanding the inner-region expansion Eq. \parref{eqn:method:ansatz} for $\brp\to\infty$. For the 1D vanishing nucleus case, expanding Eq. \parref{eqn:method:ansatz} for $\abs{\bar x}\to\infty$ yields
\ben
\begin{split}
&\psi^\text{outer}(s,\bar x)\stackrel{\abs{\bar x}\to\infty}{\sim} (1-\sqrt{2}s\abs{\bar x}+s^2{\bar x}^2+\cdots)\\
&\quad+\frac{is^2}{2}(1-\sqrt{2}s\abs{\bar x}+\cdots)+\sqrt{\frac{i}{2\pi}}^*\frac{s\exp(i{\bar x}^2)}{{\bar x}^2}+\cdots\\
&=(1-\abs{x}+\frac{x^2}{2}+\cdots)+\frac{it}{2}(1-\abs{\bar x}+\cdots)\\
&\quad+\sqrt{\frac{2i}{\pi}}^*\frac{\exp[ix^2/(2t)]}{x^2}t^{3/2}+\cdots.
\end{split}
\een
This result agrees with Eq. \parref{eqn:1dnucdisapp:outer}, except that the $\exp(-\abs{x})$ envelope of the regular terms in Eq. \parref{eqn:1dnucdisapp:outer} is expanded at $x\to0$, as the price paid for obtaining the outer-region expansion from the inner-region expansion. The same result is obtained with the stationary phase approximation\cite{BO99}.

To find the asymptotic behavior without solving TDSE, we use
the method of dominant balance\cite{BO99}.
For this case the leading-order time-non-analyticity is in
$\psi_{(1)}(\bar x)$. We use the following ansatz for $\psi_{(1)}$:
\ben
\psi_{(1)}(\bar x)=\exp[P(\bar{x})]
\label{eqn:dombal:psi1ansatz}
\een
Inserting Eq. \parref{eqn:dombal:psi1ansatz} into Eq. \parref{eqn:101229:diffeqn} yields
\ben
\label{eqn:110104:seqn}
P''(\bar{x})+[P'(\bar{x})]^2-2i\bar{x}P'(\bar{x})+2i=0
\een
One consistent balance is assuming $P''(\bar{x})\ll[P'(\bar{x})]^2$. We obtain the reduced differential equation corresponding to this balance by removing $P''(\bar x)$ from Eq. \parref {eqn:110104:seqn}:
\ben
P(\bar{x})\sim i\bar{x}^2,
\een
which is the first order in the asymptotic series of $P({\bar x}\to\infty)$, corresponding to this balance. The next order is found by inserting
\ben
\label{eqn:110104:nextorder}
P(\bar{x})\sim i\bar{x}^2+C(\bar{x})
\een
into Eq. \parref{eqn:110104:seqn}, which yields
\ben
C(\bar{x})\sim-2\ln(\bar{x})
\label{eqn:dom:C}
\een
Thus $\psi_{(1)}$ has the following asymptotic behavior from the balance $P''(\bar{x})\ll[P'(\bar{x})]^2$:
\ben
\psi_1(\bar{x})=c_1\exp[P(\bar{x})]\sim c \frac{\exp(i\bar{x}^2)}{\bar{x}^2}
\een
More terms are obtained by iteration, and inserting results of Eq. \parref{eqn:dom:C} yields $\psi_{(1)}$ as
\ben
\label{eqn:dombal:part1}
\psi_{(1)}(\bar{x})\sim c_1\frac{\exp(i\bar{x}^2)}{\bar{x}^2}\sum_{n=0}^\infty\frac{(2n+1)!!(-i)^n}{2^n}{\bar x}^{-2n}.
\een

Another consistent balance is assuming $-2i\bar{x}P'(\bar{x})\gg P''(\bar{x}),[P'(\bar{x})]^2$. The asymptotic series corresponding to this balance is
\ben
\psi_{(1)}(\bar{x})\sim c_2\bar{x}.
\label{eqn:dombal:part2}
\een
The complete asymptotic behavior is then a summation of Eq. \parref{eqn:dombal:part1} and Eq. \parref{eqn:dombal:part2}:
\ben
\label{eqn:110104:dombalresult}
\psi_{(1)}(\bar{x})\sim c_1 \frac{\exp(i\bar{x}^2)}{\bar{x}^2}\left(1-\frac{3i}{2\bar{x}^2}-\frac{15}{4\bar{x}^4}+\cdots\right)+c_2 \bar{x}.
\een

Borel summation is a method of extracting information and yields the closed-form formula of a function from its asymptotic series under certain restrictions\cite{BO99, S95}. Consider a divergent series
\ben
S(p)=\sum_{n=0}^\infty \beta_n p^n.
\een
The Borel sum of the series is defined as
\ben
S_\text{B}(p)\equiv\int_0^\infty\intd \xi\;\exp(-\xi)\phi(p\xi),
\een
in which
\ben
\phi(p)=\sum_{n=0}^\infty\frac{\beta_n p^n}{n!}.
\een

As an example, we do the Borel sum of the series in Eq. \parref{eqn:110104:dombalresult}. The original divergent series is
\ben
S(\bar{x})=\frac{\exp(i\bar{x}^2)}{\bar{x}^2}\sum_{n=0}^\infty\frac{(2n+1)!!(-i)^n}{2^n}\bar{x}^{-2n}.
\label{eqn:borel:originalseries}
\een
The Borel sum of Eq. \parref{eqn:borel:originalseries} is
\ben
\begin{split}
S_\text{B}(\bar{x})&=\frac{\exp(i\bar{x}^2)}{\bar{x}^2}\int_0^\infty\intd \xi\;\exp(-\xi)\sum_{n=0}^\infty\frac{(2n+1)!!(-i)^n\xi^n}{2^n n!\,\bar{x}^{2n}}\\
&=\frac{\exp(i\bar{x}^2)}{\bar{x}^2}\int_0^\infty\intd \xi\;\exp(-\xi)\frac{1}{(1+i\xi/\bar{x}^2)^{3/2}}\\
&=-2i\exp(i\bar{x}^2)+2\sqrt{i\pi}\bar{x}\erfc\left(\sqrt{i}^*\bar{x}\right),
\end{split}
\een
which is the exact form of $\psi_{(1)}$ as in Eq. \parref{eqn:101229:exactsoln}:
\ben
\psi_{(1)}(\bar{x})=c_2\bar{x}+c_1\left[-2i\exp(i\bar{x}^2)+2\sqrt{i\pi}\bar{x}\erfc\left(\sqrt{i}^*\bar{x}\right)\right],
\een
and $c_1$, $c_2$ are obtained by matching with Eq. \parref{eqn:initcond}.
\ben
c_1=\sqrt{\frac{i}{\pi}}^*,\quad c_2=-\sqrt{2}.
\een
This result agrees with Eq. \parref{eqn:101229:exactsoln}, so the method worked.

\subsection{Suddenly switched electric field}
\label{sect:app:1dHinEfield}
Next we apply the method on a more complicated 1D 1-electron case. Consider a system with the following potential:
\ben
V(x,t)=-\delta(x)+\calE x\theta(t).
\een
The initial state is Eq. \parref{eqn:intro:psiinit}, the ground state of the `1D hydrogen'. The system stays in that state for $t<0$, and a static linear electric field with field strength $\calE$ is turned on at $t=0$. Though we cannot obtain the full analytic wavefunction for this system, the 1st order perturbative wavefunction(sans $\calE$) $\psi^{<1>}$\cite{E06} is exactly solvable and is given in Sect. \ref{sect:freq:nonanalyticity}.
Its outer expansion to the leading time-non-analytic order is
\begin{multline}
\psi^{<1>}(x,t)\stackrel{t\to0_+}{\sim} \left\{-ixt+\frac{x-\sgn(x)}{2}t^2\right.\\
\left.+\frac{i[x-\sgn(x)]}{8}t^3-\frac{x-\sgn(x)}{48}t^4\right\}\exp(-\abs{x})\\
-4\sqrt{\frac{2i}{\pi}}\frac{\exp[ix^2/(2t)]}{x^5}t^{9/2}+O(t^5),
\label{eqn:1dHinEfield:exactshortt}
\end{multline}
Below we show that the $s$-expansion method reproduces the $t^{9/2}$ term in Eq. \parref{eqn:1dHinEfield:exactshortt}.

Although $V$ has no explicit time-dependence for $t>0$, the potential in the reduced variables has explicit $s$-dependence, which is
\ben
\label{eqn:1dHinEfield:pot}
V(x,t>0)=-\delta(x)+\calE x=-\frac{\delta(\bar{x})}{s\sqrt{2}}+\calE s\bar{x}\sqrt{2}.
\een
With the $s$-dependent potential Eq. \parref{eqn:1dHinEfield:pot}, the differential equations Eq. \parref{eqn:method:diffeqn} become a system of inhomogeneous differential equations:
\ben
\label{eqn:101012:1dHinEfieldDiffEqn}
\psi_{(n)}''-2i\bar{x}\psi_{(n)}'+2in\psi_{(n)}+2\sqrt{2}\delta(\bar{x})\psi_{(n-1)}-4\sqrt{2}\calE \bar{x}\psi_{(n-3)}=0.
\een
The boundary conditions for Eq. \parref{eqn:101012:1dHinEfieldDiffEqn} is the same as Eq. \parref{eqn:initcond}, since the initial condition of TDSE does not change from Eq. \parref{eqn:intro:psiinit}. A general formula for $\psi_{(n)}(\bar x)$ like Eq. \parref{eqn:101229:exactsoln} is not available in this case.

Converting the $t^{9/2}$ term in Eq. \parref{eqn:1dHinEfield:exactshortt} to $(s,\bar x)$ variables, we observe the leading-order time-non-analyticity occurs at fourth order in $s$, and we solve for $\psi_{(4)}(\bar x)$ for this time-non-analyticity. Since $\psi_{(4)}(\bar x)$ depends on all the previous $\psi_{(n)}(\bar x)$'s as shown in Eq. \parref{eqn:101012:1dHinEfieldDiffEqn}, we need $\psi_{(0)}(\bar x)$ to $\psi_{(3)}(\bar x)$ to solve for $\psi_{(4)}$.

In this case, $\psi_{(0)}(\bar x)$ and $\psi_{(3)}(\bar x)$ can be obtained easily from Eq. \parref{eqn:101012:1dHinEfieldDiffEqn}. For a more complicated system, there may be more such extra work to do before reaching the leading-order time-non-analyticity, and it is cumbersome having to solve for the first few $\psi_{(n)}$'s which are analytic in time. We observe that though the $t$-TE wavefunction does not have the correct short-time behavior, it can be used to facilitate the process of obtaining the leading-order time-non-analyticity in $\psi(x,t)$, as described below.

$\psi^\text{TE}(x,t)$ of this system is
\begin{multline}
\psi^\text{TE}(x,t)=\psi_0(x)\left[1-it\left(-\frac{1}{2}+\calE x\right)\right.\\
\left.-\frac{t^2}{2}\left(\frac{1}{4}+\calE \sgn(x)-\calE x+\calE^2x^2\right)\right]+O(t^3).
\label{eqn:1dHinEfield:psiTE}
\end{multline}
Converting Eq. \parref{eqn:1dHinEfield:psiTE} to $(s,\bar x)$ variables and collecting the $s^n$ terms gives a set of $\psi^\text{TE}_{(n)}(\bar x)$. $\psi_{(n)}^\text{TE}$ for $n=0\sim 4$ are listed below:
\ben
\begin{split}
\psi_{(n=0,1,2)}^\text{TE}(\bar x)&=\psi_{(n),\calE=0}^\text{TE}(\bar x)\\
\psi_{(3)}^\text{TE}(\bar x)&=\psi_{(3),\calE=0}^\text{TE}(\bar x)-i\sqrt{2\calE}{\bar x},\\
\psi_{(4)}^\text{TE}(\bar x)&=\psi_{(4),\calE=0}^\text{TE}(\bar x)-\calE\sgn(\bar x)\left(\frac{1}{2}-2i{\bar x}^2\right),
\end{split}
\een
with $\psi_{(n),\calE=0}^\text{TE}(\bar x)=(n!)^{-1}\partial^n[\psi_0(s,\bar x)\exp(is^2/2)]/\partial s^n|_{s=0}$.

$\psi_{(0)}^\text{TE}(\bar x)$ to $\psi_{(3)}^\text{TE}(\bar x)$ satisfy both the differential equations Eq. \parref{eqn:101012:1dHinEfieldDiffEqn} and the boundary conditions Eq. \parref{eqn:initcond}, which is expected since the outer-expansion Eq. \parref{eqn:1dHinEfield:exactshortt} suggests the leading-order time-non-analyticity does not occur until $\psi_{(4)}$. Inserting $\psi_{(4)}^\text{TE}(\bar x)$ into the left hand side of Eq. \parref{eqn:101012:1dHinEfieldDiffEqn} yields $-\calE\delta'(\bar{x})$, showing that $\psi_{(4)}^\text{TE}(\bar x)$ does not satisfy the differential equation. Then we only need to solve the differential equations starting from $\psi_{(4)}(\bar x)$.

We define the difference between $\psi_{(4)}$ and $\psi_{(4)}^\text{TE}$ as
\ben
\Delta(\bar x)=\psi_{(4)}(\bar x)-\psi_{(4)}^\text{TE}(\bar x).
\een
Then Eq. \parref{eqn:101012:1dHinEfieldDiffEqn} in terms of $\Delta$ becomes
\ben
\Delta''-2i\bar{x}\Delta'+8i\Delta-\calE\delta'(\bar{x})=0.
\label{eqn:1dHinEfield}
\een
We obtain the complete asymptotic expansion of the general solution $\Delta_g$ for $\bar x\to\infty$ by the method of dominant balance(as described in Sect. \ref{sect:app:1ddisappnuc}):
\begin{multline}
\Delta_g(\bar x)\sim c_1\left(\bar{x}^4+3i\bar{x}^2-\frac{3}{4}\right)\\
+c_2\frac{\exp(i\bar{x}^2)}{\bar{x}^5}\left[1+\frac{1}{3\bar{x}^2}\sum_{m=0}^\infty\frac{(2m+6)!(-i)^{m+1}}{(m+1)!2^{2m+5}}\bar{x}^{-2m}\right],
\label{eqn:1dHinEfield:asympfull}
\end{multline}
in which $c_1$ and $c_2$ are coefficients to be determined later. We apply a Borel summation to Eq. \parref{eqn:1dHinEfield:asympfull}, which yields the exact formula for $\Delta_g(\bar x)$:
\begin{multline}
\label{eqn:110111:1dHinEfield}
\Delta_g(\bar{x})=c_1(\bar{x}^4+3i\bar{x}^2-3/4)+c_2\left[-\frac{1}{3}\exp(i\bar{x}^2)\bar{x}(5i+2\bar{x}^2)\right.\\
\left.+\frac{1-i}{6}\sqrt{\frac{\pi}{2}}(-3+12i\bar{x}^2+4\bar{x}^4)\erfc\left(\sqrt{i}^*\bar{x}\right)\right]\Bigg\}.
\end{multline}
The coefficients $c_1$ and $c_2$ are determined using the boundary conditions Eq. \parref{eqn:initcond}, yielding
\begin{multline}
\psi_{(4)}(\bar x)=\frac{1}{6}\bar{x}^4+\frac{1}{2}i\bar{x}^2-\frac{1}{8}\\
-\calE\left\{\frac{2}{3}\bar{x}^4\sgn(\bar{x})+\frac{\sqrt{i}}{3\sqrt{\pi}}\exp(i\bar{x}^2)\bar{x}(5i+2\bar{x}^2)\right.\\
\left.+\left(\frac{2}{3}\bar{x}^4+2i\bar{x}^2-\frac{1}{2}\right)\erf(\sqrt{i}^*\bar{x})\right\}.
\label{eqn:1DHpsi4}
\end{multline}
We obtain the leading time-non-analytic term in the outer-region expansion similarly as in Sect. \ref{sect:app:1ddisappnuc}, which is verified by Eq. \parref{eqn:1dHinEfield:exactshortt}.

Unlike Eq. \parref{eqn:1dHinEfield:exactshortt}, no expansion in powers of $\calE$ was needed. Eq. \parref{eqn:101012:1dHinEfieldDiffEqn} shows that $\psi_{(6)}(\bar x)$ contains the first $\calE^2$ term, and $\psi_{(9)}(\bar x)$ the first $\calE^3$ term.

\subsection{Time-varying nuclear charge}
\label{sect:app:restriction}
Here we discuss a more subtle restriction of the method. We study a 1D system with Eq. \parref{eqn:intro:psiinit} as the initial wavefunction, and with the following potential:
\ben
V(x,t)=-[1+\epsilon\theta(t)]\delta(x).
\een
In this system, the strength of the $\delta$-well changes at $t=0$, causing the shape of the cusp at $x=0$ to change.  The analytic form of the exact wavefunction can be written out, and the exact leading order time-non-analytic term is
\ben
\label{eqn:101223:expint}
-\sqrt{\frac{i}{2\pi}}^*\epsilon[2+(1+\epsilon)\abs{x}]e^{ix^2/(2t)}x^{-2}t^{3/2}.
\een
However, that derived with the method in Sect. \ref{sect:method} is
\ben
-\sqrt{\frac{2i}{\pi}}^*\epsilon e^{ix^2/(2t)}x^{-2}t^{3/2},
\een
which is only a part of Eq. \parref{eqn:101223:expint}. The reason for this discrepancy is that there are two time-scales in the short-time behavior of this system, one is determined by the cusp in the initial wavefunction, and the other one is determined by the $\delta$-well whose strength has changed. The simple boundary-layer analysis in Sect. \ref{sect:intro} does not apply here, as the boundary layer structure is too complicated here.

\section{Applications in 3D}
The essential methodology remains the same when turning to 3D, but the equations
become substantially more complex. For brevity, \emph{we normalize 3D wavefunctions to $\pi$ instead of $1$}.

\subsection{Vanishing nucleus}
\label{sect:3D:vanishingnucleus}
One point needs to be changed for the $s$-expansion method in 3D. Consider a system whose initial wavefunction equals the ground-state wavefunction of the hydrogen atom:
\ben
\label{eqn:3d:initwf}
\psi_0(\vect{r})=\exp(-r).
\een
Free-propagation of this wavefunction yields a similar situation as in the 1D vanishing nucleus case, as the system is effectively 1D due to the spherical symmetry. By expanding the initial wavefunction Eq. \parref{eqn:3d:initwf} as
\ben
\psi(s,\bar{r})\stackrel{s\to0_+}{\sim}1-s\sqrt{2}\bar{r}+s^2\bar{r}^2+\cdots,
\label{eqn:3d:bc}
\een
We only obtain one boundary condition($\psi_{(n)}(\bar r\to\infty)$) for Eq. \parref{eqn:method:diffeqn} instead of two boundary conditions as in 1D cases($\psi_{(n)}(\bar x\to\pm\infty)$). Eq. \parref{eqn:method:diffeqn} requires another boundary condition to be well-defined, and it is related to how $t$-TE behaves in 3D cases. For the 3D vanishing nucleus case, the $t$-TE wavefunction is
\ben
\begin{split}
\psi^\text{TE}(\vect{r},t)&=\exp(-r+it/2)\left(1-\frac{it}{r}\right)\\
&=1-\frac{i+2\bar{r}^2}{\sqrt{2}\bar{r}}s+\frac{3i+2\bar{r}^2}{2}s^2+O(s^3).
\end{split}
\een
Unlike in the 1D examples, all $\psi_{(n)}^\text{TE}(\brp)$ satisfy Eq. \parref{eqn:method:diffeqn}, but $\psi_{(1)}^\text{TE}(\brp)$ diverges at $\bar r=0$ for any non-zero time. Thus the other boundary condition for Eq. \parref{eqn:method:diffeqn} is $\psi_{(n)}(\brp)$ must be regular at $\bar r=0$.

\subsection{Suddenly switched electric field}
\label{sect:app:3dHinEfield}
We discussed 3D systems in our previous paper\cite{YMB12}. Here we provide a more detailed derivation for 3D hydrogen atom in a turned-on static electric field. Aside from the dimensionality change, the main change from 1D cases to 3D cases is that the Coulomb potential replaces the $\delta$-function potential as the singular potential. Unlike the $\delta$-function potential, the Coulomb potential is long-ranged, which makes 3D wavefunctions more complicated than their 1D counterparts.

The system has the following potential:
\ben
V(\vect{r},t)=-\frac{1}{r}+\calE z\theta(t).
\een
One can easily check with perturbation theory that the $t$-TE wavefunction of this system does not have a convergent norm, and thus it must have time-non-analyticities. Define reduced variables:
\ben
s=\sqrt{t},\,\bar r=\frac{r}{\sqrt{2t}},\,\bar z=\frac{z}{\sqrt{2t}}.
\een
The external potential in these reduced variables is
\ben
V(\br,t>0)=-\frac{1}{\sqrt{2}s \bar{r}}+\calE\sqrt{2}s \bar{z}.
\een
Inserting the wavefunction ansatz Eq. \parref{eqn:method:ansatz} into Eq. \parref{eqn:method:diffeqn} yields
\ben
\label{eqn:3dHinEfield:diffeqn}
({\cal L}+2in)\psi_{(n)}+\frac{2\sqrt{2}}{\bar{r}}\psi_{(n-1)}-4\sqrt{2}\calE \bar{z}\psi_{(n-3)}=0,
\een
where
\ben
{\cal L}=\frac{\partial^2}{\partial \bar{r}^2}+\frac{\partial^2}{\partial \bar{z}^2}+\left(1+{\bar z}\frac{\partial}{\partial{\bar z}}\right)\frac{2}{\bar r}\frac{\partial}{\partial{\bar r}}-2i\left({\bar r}\frac{\partial}{\partial{\bar r}}+{\bar z}\frac{\partial}{\partial{\bar z}}\right).
\een
For $\calE=0$, the TE is simple, and 
\ben
\psi_{(n),\calE=0}^\text{TE}(\brp)=(n!)^{-1}\partial^n[\psi_0(s,\brp)\exp(is^2/2)]/\partial s^n|_{s=0}.
\een
In the presence of the electric field, 
\ben
\psi_{(n)}^\text{TE}(\brp)=\psi_{(n),\calE=0}^\text{TE}(\brp)+i\calE{\bar z}f_{(n)}({\bar r}),\quad n\le 4,
\een
with
\ben
f_{(n\le 2)}=0,\, f_{(3)}=-\sqrt{2},\, f_{(4)}({\bar r})=1/(12{\bar r}^3)+1/(2{\bar r})+2{\bar r}.
\een
The $\psi^\text{TE}_{(n)}(\brp)$'s before the occurring of the leading-order time-non-analyticity are identical to $\psi_{(n)}(\brp)$, and we only need to solve for $\psi_{(n)}(\brp)$ if $\psi^\text{TE}_{(n)}(\brp)$ fails to satisfy the differential equation Eq. \parref{eqn:3dHinEfield:diffeqn} and the boundary conditions.
Since $\psi_{(4)}^\text{TE}(\brp)$ diverges as $\bar{r}\to0$, the leading-order time-non-analyticity is in $\psi_{(4)}(\brp)$.

As before, we use the method of dominant balance and Borel summation to solve for $\Delta(\brp)=\psi_{(4)}(\brp)-\psi_{(4)}^\text{TE}(\brp)$. Since $\psi_{(4)}^\text{TE}$ satisfies Eq. \parref{eqn:3dHinEfield:diffeqn}, the equation can be rewritten as
\ben
\label{eqn:3dHinEfield:DeltaEqn}
\left({\cal L}+8i\right)\Delta=0,
\een
As $\bar{r}\to0$, the divergence in $\psi_{(4)}^\text{TE}$ is proportional to $\bar z$, and $\Delta(\brp)$ must cancel this divergence to satisfy the boundary conditions. Therefore $\Delta(\brp)$ has the following form:
\ben
\Delta(\bar{r},\bar{z})=g(\bar{r})\bar{z}.
\een
The method of dominant balance(Sect. \ref{sect:app:1ddisappnuc}) yields the entire asymptotic expansion of $g(\brp)$:
\begin{multline}
g(\brp)=c_1\left(\bar{r}^3+\frac{9i\bar{r}}{2}-\frac{9}{4\bar{r}}+\frac{3i}{8\bar{r}^3}\right)\\
+c_2\frac{\exp(i\bar{r}^2)}{\bar{r}^8}\left[1+\frac{1}{9\bar{r}^2}\sum_{m=0}^\infty\frac{(-i)^{m+1}(m+4)(2m+6)!}{(m+1)!2^{2m+5}\bar{r}^{2m}}\right].
\end{multline}
Performing the Borel sum, we find
\begin{multline}
g(\brp)=c_1\left(\bar{r}^3+\frac{9i\bar{r}}{2}-\frac{9}{4\bar{r}}+\frac{3i}{8\bar{r}^3}\right)\\
+c_2\frac{\sqrt{2i}}{72\bar{r}^3}\bigg[2\sqrt{2i}\exp(i\bar{r}^2)\bar{r}(-3+16i\bar{r}^2+4\bar{r}^4)\\
-\sqrt{2\pi}(3i-18\bar{r}^2+36i\bar{r}^4+8\bar{r}^6)\erfc\left(\sqrt{i}^*\bar{r}\right)\bigg].
\label{eqn:3dHinEfield:Delta}
\end{multline}
The coefficients $c_1$ and $c_2$ are determined by the boundary conditions as ${\bar r}\to0$ and ${\bar r}\to\infty$, yielding
\ben
c_1=0,\quad c_2=-\sqrt{\frac{i}{\pi}}^*\calE.
\een
Expanding $s^4\psi_{(4)}(\bar r)$ for $\bar r\to\infty$ yields the outer-region expansion:
\ben
\psi^\text{outer}(\vect{r},t)\stackrel{\bar r\to\infty}{\sim}\cdots-\frac{8\sqrt{2i}^*\calE z}{\sqrt{\pi} r^8}\exp\left(\frac{ir^2}{2t}\right)t^{11/2}.
\label{eqn:3dHinEfield:outerfromlinearresp}
\een

Although $\psi_{(4)}(s,\brp)$ contains the leading order time-non-analyticity in the wavefunction, knowing it is insufficient\cite{YMB12} to derive the correct coefficient of the leading half-power in the TD dipole moment (Sec.  \ref{sect:app:mufromalpha}). Due to the coupling between $\br$ and $t$ in the wavefunction, higher order terms in the $s$-expansion can contribute to integrated properties such as the TD dipole moment. We have evidence that both $\psi_{(4)}$ and $\psi_{(5)}$ contribute, but we have been unable to find a closed-form expression for $\psi_{(5)}$.

Finally, we obtain the leading-order time-non-analytic term in the outer-region expansion of this system by applying the stationary phase approximation (as in Sect. \ref{sect:innerandouter}) to 1st order in $\delta V^{<1>}(\br',t')=z'\theta(t')$. The change in the wavefunction(sans $\calE$) is
\begin{multline}
\psi^{<1>}(\br,t)=i\int\intd^3 r'\;G^{<1>}(\br,\br',t)\psi^{<0>}(\br',0)
\label{eqn:1stpertalternativeform}
\end{multline}
in which $\psi^{<0>}$ is the ground-state wavefunction of 3D hydrogen, and
\begin{multline}
G^{<1>}(\br,\br',t)=\int\intd^3r''\int_0^{t}\intd t''\;G^{<0>}(\br,\br'',t-t'')\\
\times\delta V^{<1>}(\br'',t'')G^{<0>}(\br'',\br',t''),
\end{multline}
is the 1st order change of the Green's function, with
\ben
G^{<0>}(\br,\br',t)=-i\sum_n\exp(-i\epsilon_nt)\psi_n(\br)\psi_n^*(\br'),
\label{eqn:G0Lehmann}
\een
where $\psi_n$ and $\epsilon_n$ are atomic orbitals and orbital energies of the 3D hydrogen atom respectively. Only the well-known\cite{F06} unbound $p$ orbitals in the sum of Eq. \parref{eqn:G0Lehmann} contribute to the time-non-analyticity in $\psi^{<1>}$, and the sum in Eq. \parref{eqn:G0Lehmann} becomes an integration for unbound orbitals. Applying the stationary phase approximation(as in Sect. \ref{sect:innerandouter}) to this integration, we obtain the leading non-analytic short-time behavior shown in Eq. \parref{eqn:3dHinEfield:outerfromlinearresp}.

\section{Dipole moments}
\label{sect:app:mufromalpha}
Our results so far, using the $s$-expansion, have been for the time-dependent wavefunction. There is no simple result for
its short-time dependence, due to the existence of distinct expansions in the inner
and outer regions.  However, expectation values over wavefunctions {\em do} have well-defined
expansions for small times, although more complex than a simple Taylor expansion.

In the present section, we extract results for dipole moments induced by
turning on an electric field, in both 1D and 3D.  Note that our $s$-expansion
is not a perturbation expansion in the applied field, but rather demonstrates
that the leading corrections to the wavefunction for short-time behavior are linear in the
applied field.  On the other hand, we deduce dipole
moments only within linear-response theory, but we find consistent results, as shown below.

\subsection{Linear response theory}
The linear(1st order) change in the density $\delta n(\vect{r},t)$ is described by the linear response function $\chi$ and its Fourier transform $\tilde{\chi}$:
\ben
\begin{split}
\chi(\vect{r},\vect{r}',t-t')&=\frac{\delta n(\vect{r},t)}{\delta v(\vect{r}',t')},\\
\tilde{\chi}(\vect{r},\vect{r}',\omega)&=\int_{-\infty}^{\infty}d\tau\;\chi(\vect{r},\vect{r}',\tau)e^{i\omega\tau}.
\end{split}
\label{eqn:chidef}
\een
For one particle systems, there is a simple relation between $\tilde{\chi}$ and the frequency-domain Green's function $\tilde{G}$\cite{MWB03}:
\ben
\chi(\vect{r},\vect{r}',\omega)=\sqrt{n(\vect{r})n(\vect{r}')}\left[\tilde{G}(\vect{r},\vect{r}',\omega+\epsilon_0)+\tilde{G}^*(\vect{r},\vect{r}',\epsilon_0-\omega)\right],
\label{eqn:chifromG}
\een
where $\epsilon_0$ is the ground-state energy of the system. Aside from the definition Eq. \parref{eqn:chidef}, the linear response function can also be expressed in the Lehmann representation:
\begin{multline}
\tilde{\chi}(\vect{r},\vect{r}',\omega)=\lim_{\eta\to0_+}\sum_j\left\{\frac{\matelem{\Psi_0}{\hat{n}(\vect{r})}{\Psi_j}\matelem{\Psi_j}{\hat{n}(\vect{r}')}{\Psi_0}}{\omega-\omega_j+i\eta}\right.\\
\left.+\frac{\matelem{\Psi_0}{\hat{n}(\vect{r}')}{\Psi_j}\matelem{\Psi_j}{\hat{n}(\vect{r})}{\Psi_0}}{-\omega-\omega_j-i\eta}\right\},
\label{eqn:theory:lehmann}
\end{multline}
where $\Psi$ is the many-body eigen wavefunction of the corresponding system labeled with $j$, and $\omega_j=\epsilon_j-\epsilon_0$ is the transition frequency between state $j$ and the ground state.
For a system with only a discrete spectrum, one can take $\omega\to\infty$ for each separate term, yielding $O(\omega^{-2})$ high-frequency behavior\cite{L01}. But when the system has a continuum, the sum must be performed before taking the $\omega\to\infty$ limit, and this produces fractional decay. 

Consider a perturbation potential $\calE x\theta(t)$. The 1st order dipole moment(sans $\calE$) is
\ben
\mu_x^{<1>}(t)=\int_0^t\intd t'\int\intd^3r\int\intd^3r'\; x x'\chi(\vect{r},\vect{r}',t-t'),
\label{eqn:1dHinEfield:mudef}
\een
and its transform, the polarizability in $x$ direction is
\ben
\alpha_{xx}(\omega)=\int\intd^3r\int\intd^3r'\; x x'\tilde{\chi}(\vect{r},\vect{r}',\omega).
\een
The subscripts denotes the direction on which these observables are measured. $\alpha_{xx}(\omega)$ and $\mu_x^{<1>}(t)$ are related by Fourier transform, and the high-frequency behavior of $\alpha_{xx}(\omega)$ depends on the short-time behavior of $\mu_x^{<1>}(t)$.

\subsection{Known results}
The high-frequency part
of the photoabsorption cross-section of all atoms decays as $\omega^{-7/2}$\cite{BS57,FC68,YFB09},
which means that the $\Im\alpha$ decays as $\omega^{-9/2}$ ($\Im$ denotes the imaginary part).   For hydrogen-like atoms,
$\Im[\alpha(\omega\to\infty)]$ is\cite{BS57}
\ben
\Im[\alpha(\omega)]\stackrel{\omega\to\infty}{\sim}\frac{4\sqrt{2}}{3\omega^{9/2}},
\label{eqn:imalphaasym}
\een
where $\alpha$ is a spherical average. Thus
\ben
\mu^{<1>}(t)\stackrel{t\to0_+}{\sim}\frac{2}{\pi}\int_0^t\intd t'\int_{\omega_c}^\infty\intd\omega\;\Im[\alpha(\omega)]\sin[\omega(t-t')],
\label{eqn:imalphatomu}
\een
where $\omega_c\gg 1$ is a cut-off, yielding
\begin{multline}
\mu^{<1>}(t)\stackrel{t\to0_+}{\sim}\int_0^t\intd t'\;\bigg\{\frac{16\sqrt{2}(t-t')}{15\pi\omega_c^{5/2}}-\frac{8\sqrt{2}(t-t')^3}{9\pi\omega_c^{1/2}}\\
+\frac{128(t-t')^{7/2}}{315\sqrt{\pi}}\bigg\}+\cdots,
\label{eqn:muleadinghalfpower}
\end{multline}
As $\omega_c\to\infty$, we find the leading time-non-analytic term in $\mu^{<1>}(t\to0_+)$:
\ben
\mu^{<1>}(t)\stackrel{t\to0_+}{\sim}\cdots+\frac{256}{2835\sqrt{\pi}}t^{9/2}+\cdots.
\een

\subsection{Origins of non-analyticity and relation to time-dependence}
\label{sect:freq:nonanalyticity}
To trace clearly the origin of these non-analytic behaviors, we begin with the
simplest case, a free particle in 1D.  The green's function is simply
\ben
\tilde{G}^\text{free}(x,x',\omega)=\frac{\exp(iku)}{ik},
\een
where $u=|x-x'|$ and $k=\sqrt{2\omega}$.  Insertion into Eq. \parref{eqn:chifromG} yields $\tilde{\chi}$:
\ben
\tilde{\chi}^\text{free}(x,x',\omega)=-\frac{\exp(-ku)+i\exp(iku)}{k}.
\een
Even for a free particle, there are non-analytic behaviors in the frequency-dependent
response due to the continuum, which are not apparent in the Lehmann representation Eq. \parref{eqn:theory:lehmann}.

Our next example is the 1D H atom.  Here
\begin{multline}
G^\text{1DH}(x,x',\tau)=-i\sqrt{\frac{1}{2\pi i\tau}}\exp\left[\frac{iu^2}{2\tau}\right]\\
-\frac{i}{2}\exp\left(\frac{i\tau}{2}-X\right)\erfc\left(\frac{X}{\sqrt{2i\tau}}-\sqrt{\frac{i\tau}{2}}\right),
\end{multline}
where $X=\abs{x}+\abs{x'}$, leading to a response function of the form:
\begin{multline}
\tilde{\chi}^\text{1DH}(x,x',\omega)=-i\exp(-X)\left[\frac{\exp(iu\kappa_+)}{\kappa_+}-\frac{\exp(iX\kappa_+)}{i\kappa_+^2+\kappa_+}\right.\\
\left.-\frac{\exp(-iu\kappa_-^*)}{\kappa_-^*}-\frac{\exp(-iX\kappa_-^*)}{i\kappa_-^2-\kappa_-^*}\right],
\label{eqn:chi1DH}
\end{multline}
where $\kappa_\pm=\sqrt{\pm 2\omega-1}$. Eq. \parref{eqn:chi1DH} clearly has non-analytic behavior for large $\omega$.  

For a 1D H atom in a turned-on linear electric field, we can explicitly calculate the 1st order perturbative wavefunction (sans $\calE$):
\begin{multline}
\psi^{<1>}(x,t)=\sqrt{\frac{i}{\pi}}\exp[i{\bar x}^2]{\bar x}t+\frac{\exp(it/2)}{2}\Big\{h_+(x,t)+h_-(x,t)\\
+\left[(\abs{x}+t^2)\sinh(x)-x(\abs{x}+2it)\cosh(x)\right]\Big\},
\end{multline}
where $h_\pm(x,t)=-\exp(\pm x)(x\mp y_\pm^2) \erf[\sqrt{i/(2t)}^{\,*}y_\pm]/2$, with $y_\pm=\pm x+it$, and ${\bar x}=x/\sqrt{2t}$ as before.

The induced first-order time-dependent dipole moment(sans $\calE$) $\mu^{<1>}$ is related to $\psi^{<1>}$ by
\ben
\mu^{<1>}(t)=2\Re\matelem{\psi^{<0>}}{x}{\psi^{<1>}},
\label{eqn:mufrompsi}
\een
($\Re$ denotes the real part)
so $\mu^{<1>}$ of this system is then
\begin{multline}
\mu^{<1>}(t)=-\frac{t^2}{12}(t^2+6)+\frac{\sqrt{t}\cos(t/2)}{12\sqrt{\pi}}(t^3-3t^2+7t+15)\\
-\frac{\sqrt{t}\sin(t/2)}{12\sqrt{\pi}}(t^3+3t^2+7t-15)\\
+2\Re\left\{\frac{1}{24}(t^4-4it^3+6t^2+12it-15)\erf\left(\sqrt{\frac{it}{2}}\right)\right\}.
\end{multline}
The leading short-time behavior is
\ben
\mu^{<1>}(t)\stackrel{t\to0_+}{\sim}-\frac{t^2}{2}+\frac{32}{105\sqrt{\pi}}t^{7/2}-\frac{t^4}{12}+O(t^{9/2}).
\label{eqn:1dHinEfield:muasymp}
\een

To see the connection with the $s$-expansion in this case, we note simply
that the 4-th order contribution Eq. \parref{eqn:1DHpsi4}, inserted in Eq. \parref{eqn:mufrompsi}, recovers the same
leading non-analytic behavior.  Thus, here, the leading-order
non-analyticity in the wavefunction is sufficient to determine the
leading-order non-analyticity in the dipole moment, at least
to first-order in the external electric field.

\section{More general potentials}
Here we explore what happens for other potentials.

\subsection{Different spatial dependence}
\label{sect:app:tnonanalyfromcusp}
We provide a heuristic demonstration that the time-non-analyticities originate from the specific form of the TDSE, and show that the time-non-analyticity of the time-dependent wavefunction is determined by the space-non-analyticity of the initial wavefunction.

Consider a perturbed 1D one-electron model system described by the following potential:
\ben
V(x,t)=V_0(x)+\calE x^n\theta(t).
\label{eqn:transform:modelsys}
\een
The structure of the problem is exposed by taking a space-Fourier transform and a time-Laplace transform of the TDSE:
\begin{multline}
\frac{k^2}{2}\tilde{\Psi}(k,\nu)+V_0(k)\ast\tilde{\Psi}(k,\nu)+\calE i^n\tilde{\Psi}^{(n)}(k,\nu)\\
-i\nu\tilde{\Psi}(k,\nu)+i\Psi_0(k)=0,
\end{multline}
where $\tilde{\Psi}(k,\nu)$ is the time-Laplace and spatial Fourier transform of $\psi(x,t)$, $\Psi_0(k)$ is the spatial Fourier transform of $\psi_0(x)$, $\ast$ denotes convolution, and the superscript $(n)$ denotes $n$-th order derivative with respect to $k$. For analytic $V_0(x)$, the $V_0(k)\ast\tilde{\Psi}(k,\nu)$ term is composed of derivatives of $\tilde{\Psi}(k,\nu)$. Our goal is to find out the short-time behavior of the time-dependent wavefunction. Dividing through by $\nu$ and taking $\nu$ large, the highest derivative is multiplied by a small parameter, and the solution of such an equation has a so-called boundary layer behavior\cite{BO99,W10}. This means the solution changes its behavior rapidly in a narrow region whose thickness is determined by the small parameter. Using boundary layer theory, we obtain a very crude estimate of the outer-region expansion of the time-dependent wavefunction by dropping all derivative terms\cite{BO99}:
\ben
\frac{k^2}{2}\tilde{\Psi}(k,\nu)-i\nu\tilde{\Psi}(k,\nu)+i\Psi_0(k)=0,
\een
yielding
\ben
\tilde{\Psi}(k,\nu)=-\frac{2i\Psi_0(k)}{k^2-2i\nu}.
\label{eqn:transform:psinuk}
\een
This specific pole structure is due to the specific form of the TDSE, that of a 2nd order differential equation in space, but a 1st order differential equation in time. This pole structure generates the time-non-analyticities shown in the previous examples. One recognizes this by doing the inverse Laplace/Fourier transform of the pole:
\ben
\psi(x,t)=\sqrt{\frac{i}{\pi t}}^{\,*}\exp\left(\frac{ix^2}{2t}\right),\quad\text{for }\Psi_0(k)=1.
\een

Though the form of the TDSE implies time-non-analyticities, such non-analyticities do not show up in every system. If the initial wavefunction is analytic in space, then the time-dependent wavefunction of the system described by Eq. \parref{eqn:transform:modelsys} is analytic in time; if the initial wavefunction has cusps, the time-dependent wavefunction is not time-analytic, and the time-non-analyticities have the form $t^{n/2}$ and $\exp[ix^2/(2t)]$. 

The inverse Laplace transform of Eq. \parref{eqn:transform:psinuk} is
\ben
\Psi(k,t)\sim-\exp(-ik^2t/2)\Psi_0(k).
\label{eqn:transform:inverlaplace}
\een
The outer-region asymptotic behavior of $\psi(x,t\to0_+)$ is obtained from the inverse Fourier transform of Eq. \parref{eqn:transform:inverlaplace}:
\ben
\psi(x,t)\sim-\frac{\exp[ix^2/(2t)]}{\sqrt{it}}\ast\Psi_0(x),
\label{eqn:transform:backtopsi}
\een
in which $\ast$ denotes convolution. If $\psi(x,t)$ is space-analytic, it equals its Taylor expansion:
\ben
\psi_0(x)=\sum_{j=0}^\infty\frac{\psi^{(j)}_0(0)}{j!}x^j,
\label{eqn:transform:psiTE}
\een
where $\psi^{(j)}_0$ here denote $j$-th order space derivative of $\psi_0$.
Then the convolution in Eq. \parref{eqn:transform:backtopsi} can be evaluated term by term, with the $j$-th term being proportional to
\ben
t^{(j-l)/2}\,_1\mathrm{F}_1\left(\frac{l-j}{2};\frac{1}{2}+l;\frac{ix^2}{2t}\right),\quad l=\frac{1-(-1)^j}{2},
\label{eq:transform:analytic}
\een
where $_1\mathrm{F}_1$ is Kummer's confluent hypergeometric functions\cite{AS72}. The $_1\mathrm{F}_1$'s in Eq. \parref{eq:transform:analytic} are polynomials that involve only positive integer powers of $t$, so there are no time-non-analyticities starting from a space-analytic initial wavefunction for the model system Eq. \parref{eqn:transform:modelsys}.

For initial wavefunctions with cusps, we modify Eq. \parref{eqn:transform:psiTE} to be:
\ben
\psi_0'(x)=\sum_{j=0,j\ne m}^\infty\frac{\psi^{(j)}_0(0)}{j!}x^j + c x^m[\theta(x)-\theta(-x)],
\label{eqn:transform:psiTEwithCusp}
\een
which contains a derivative discontinuity (i.e., `cusp') in the $m$-th order. The convolution in Eq. \parref{eqn:transform:backtopsi} for the $\theta(x)$ part of Eq. \parref{eqn:transform:psiTEwithCusp} is proportional to
\ben
t^{m/2}\,_1\mathrm{F}_1\left(-\frac{m}{2};\frac{1}{2};\frac{ix^2}{2t}\right)+t^{(m-1)/2}\,_1\mathrm{F}_1\left(\frac{1-m}{2};\frac{3}{2};\frac{ix^2}{2t}\right).
\label{eqn:transform:analyticwithCusp}
\een
The convolution with the $\theta(-x)$ part yields a similar result. As in previous case, the $_1F_1$'s in Eq. \parref{eqn:transform:analyticwithCusp} are regular polynomials. Eq. \parref{eqn:transform:analyticwithCusp} contains $t$-half-powers for all values of $m$, and thus the initial wavefunction with cusps has time-non-analyticities in its short-time behavior for the model system Eq. \parref{eqn:transform:modelsys}.

We provide the free-propagation of a Gaussian initial wavefunction as an example in which there is no non-analytic short-time behavior starting from a smooth initial wavefunction. The initial wavefunction is
\ben
\psi_0(x)=\frac{\exp[-x^2/(2\sigma)^2]}{\pi^{1/4}\sqrt{\sigma}},
\een
in which $\sigma$ characterizes the width of the Gaussian. Combining Eq. \parref{eqn:transform:inverlaplace} and Eq. \parref{eqn:transform:backtopsi} yields
\ben
\psi(x,t)\sim-\frac{i\sqrt{2\sigma}\pi^{1/4}}{\sqrt{t-i\sigma^2}}\exp\left[\frac{ix^2}{2(t-i\sigma^2)}\right].
\label{eqn:transform:Gaussian}
\een
Eq. \parref{eqn:transform:Gaussian} has no time-non-analyticities at the initial time. The radius of convergence of the $t$-TE at the initial time is $\sigma^2$. In the limit of $\sigma\to0$, the Gaussian becomes a $\delta$-function and no longer smooth. The pole in Eq. \parref{eqn:transform:Gaussian} coincides with $t=0$, and as a consequence the radius of convergence of the $t$-TE becomes exactly zero(just as in Sect. \ref{sect:illus:radconv:1ddisappnuc}).

\subsection{Different time dependence}
\label{sect:notsuddenswitching}
Next we consider cases other than sudden switching.
For ease of discussion, we limit ourselves to 1D systems with the following time-dependent potential:
\ben
V(x,t)=-\delta(x)+V_a(x)+\calE\delta V^{<1>}(x,t),
\een
where $V_a(x)$ is an analytic potential, $\delta V^{<1>}(x,t)=x^n f(t)$, and $f(t)$ determines how the perturbation is turned on. At $t=0$, the system starts in the ground state $\psi_0(x)$ of potential $-\delta(x)+V_a(x)$, which has a cusp at $x=0$ due to the $\delta$-function part of the potential. 

To show that the information at the cusp is enough to determine the leading half-power term in time, we make a drastic approximation: the wavefunction is approximated by an envelope function for all $x\ne0$. Write
\ben
\psi_0(x)=g(ax)\sum_{j=0}^\infty d_jx^j,
\een
where $g(x)$ is some decaying envelope function, $a$ is a positive constant, and $d_j$'s are the Taylor coefficients of $\psi_0(x)/g(ax)$. Choosing $g(x)=\exp(-\abs{x})$ and applying Kato's cusp condition\cite{K57}, we obtain
\begin{multline}
\psi_0(x)=\exp(-a\abs{x})\Big\{\psi_0'(0_-)-\psi_0'(0_+)\\
+x\left[\psi_0'(0_+)+a\left(\psi_0'(0_-)-\psi_0'(0_+)\right)\right]\\
+x^2\left[a\psi_0'(0_+)+a^2\left(\psi_0'(0_-)-\psi_0'(0_+)\right)+\psi_0''(0_+)\right]+\cdots\Big\}
\end{multline}
We can still use Eq. \parref{eqn:imalphatomu} for the dipole moment even though the potential is more general, by defining $\Im[\hat{\alpha}(\omega)]$ in analog of the dynamic polarizability as
\ben
\Im[\hat{\alpha}(\omega\to\infty)]\sim\calE\pi\matelem{\psi_0}{x}{\psi_{k_\omega}}\matelem{\psi_{k_\omega}}{\delta V^{<1>}(x,\omega)}{\psi_0},
\label{eqn:conclusion:imalphahigh}
\een
where $\psi_{k_\omega}$ is the continuum wavefunction whose energy difference to the ground state is $\omega$, and $\delta V^{<1>}(x,\omega)$ is the Fourier transform of $\delta V^{<1>}(x,t)$. Inserting Eq. \parref{eqn:conclusion:imalphahigh} into Eq. \parref{eqn:imalphatomu}, we find that there is one term of the result which does not depend on the cut-off $\omega_c$ and the envelope parameter $a$, which is the term of the leading time-non-analyticity. This result does not depend on which smooth decaying envelope function is chosen for $g(x)$.

In previous examples, the time-dependent perturbation is always turned-on with $f(t)=\theta(t)$, allowing the possibility that the time-non-analyticity is related to this specific turning-on method. Here we test different turn-on functions $f(t)$. If $f(t)=\delta(t)$, we obtain the leading half-power term in $\mu^{<1>}$ as
\begin{multline}
\mu^{<1>}(t\to0_+)\sim\cdots+2[\psi_0'(0_+)-\psi_0'(0_-)]^2\\
\times\Gamma(-2-n/2)\Gamma(n+2)t^{2+n/2} 2^{-3-n/2}\\
\times[-1+(-1)^n](-1+i^n)\exp(-3in\pi/4)+\cdots.
\end{multline}
In another case, if $f(t)=t^m$, we obtain the leading half-power term as
\begin{multline}
\mu^{<1>}(t\to0_+)\sim\cdots+-2\frac{[\psi'(0_+)-\psi'(0_-)]^2\csc(n\pi/2)}{\Gamma(4+m+n/2)}\\
\times t^{3+m+n/2}2^{-3-n/2}[-1+(-1)^n](-1+i^n)\\
\times\Gamma(1+m)\Gamma(2+n)\exp(-3in\pi/4)+\cdots.
\end{multline}
It is clear that the effect of different turning-on method only changes the order of the non-analytic behavior, so the previously shown time-non-analytic behavior is not the result of the $\theta$-function turning-on. Similarly, the spatial part of the time-dependent perturbation potential also does not need to be in the form of $x^n$, and it can be easily tested that a perturbation of $\delta V^{<1>}(x,t)=\sin(kx-\omega_0t)\theta(t)$ also have time-non-analyticities in the short-time behavior of the wavefunction.

\subsection{Onset of non-analytic behavior}
\label{sect:app:numerical}
A nucleus has a finite radius, and one may argue that the failure of the $t$-TE due to cusps is artificial. To examine this effect, we provide a numerical example similar to the 1D hydrogen in a turned-on static electric field case, but with a rounded cusp. This is done by substituting the potential in Sect. \ref{sect:app:1dHinEfield} $-\delta(x)+\calE x\theta(t)$ with
\ben
V(x,t)=-\frac{\exp[-x^2/(2\sigma^2)]}{\sigma\sqrt{2\pi}}+\calE x\theta(t).
\een
We set $\calE=1$ for the numerical calculation. In the limit of $\sigma\to0$, the case in Sect. \ref{sect:app:1dHinEfield} is recovered. We solve the ground state wavefunction of this system for $t<0$ on an unevenly distributed grid, which has more points near $x=0$ to ensure the cusp-like structure in the wavefunction is well-resolved. We propagate the TD wavefunction with the $t$-TE based Crank-Nicolson method\cite{PTVF07}, and a sufficiently small time step considering the radius-of-convergence problem. We then calculate the numerical TD dipole moment $\mu(t)$, and fit $\mu(t)$ with
\ben
\mu(t)\sim c t^2+\frac{32}{105\sqrt{\pi}}t^{7/2},
\een
which are the first two terms of Eq. \parref{eqn:1dHinEfield:muasymp}. In the limit $\sigma\to0$, $c=-1/2$; with finite value of $\sigma$, we list the corresponding values of $c$ in Table \ref{table:numerical}.

\begin{table}[htbp]
\caption{Relation between the coefficient of $t^2$ in $\mu(t)$ and $\sigma$}
\begin{tabular}{|c|cccccc|}
\hline
$40\sigma$ & 1 & 1/2 & 1/4 & 1/8 & 1/16 & 1/32\\
$-1000(c+1/2)$ & 86.6 & 41.9 & 20.1 & 9.32 & 3.98 & 1.32\\
\hline
\end{tabular}
\label{table:numerical}
\end{table}

\begin{figure}[htbp]
\includegraphics[height=\columnwidth,angle=-90]{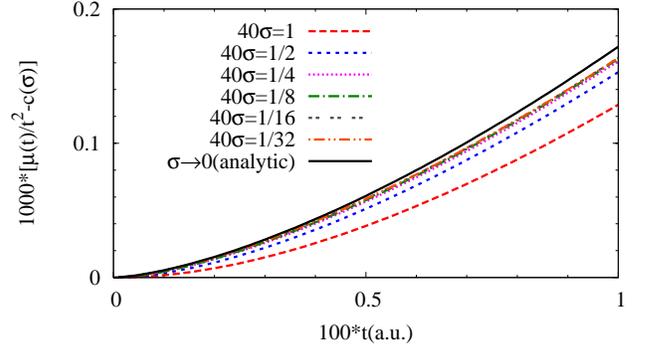}
\caption{(Color online) $\mu(t)/t^2-c(\sigma)$ for $\sigma$ listed in Table. \ref{table:numerical} and $\sigma\to0$. ($32t^{3/2}/(105\sqrt{\pi})$ is plotted for the $\sigma\to0$ curve.) The errors between the exact $\sigma\to0$ curve and the fit curves show systematic behavior.}
\label{fig:numerical}
\end{figure}

Table \ref{table:numerical} shows that although the system does not have a cusp, the TD behavior in Sec.  \ref{sect:app:1dHinEfield} heavily influences the TD behavior here. This is hardly surprising as Sec.  \ref{sect:app:1dHinEfield} correspond to the $\sigma\to0$ limit. We used the $t$-TE based propagation scheme in the numerical example, and the $t^{7/2}$-like behavior is mimicked by all the integer $t$ powers in the $t$-TE when $\sigma$ is small. Fig. \ref{fig:numerical} shows that the $t^{7/2}$ term in Eq. \parref{eqn:1dHinEfield:muasymp} and $\mu(t)-c(\sigma)t^2$ with small $\sigma$ are nearly identical. Thus the time-non-analyticity is still relevant in numerical situations. On atomic time scale, the time evolution is indistinguishable from that with a cusp.

\section{Discussion: Many-electron systems and TDDFT}
\label{sect:disscussion}
The original motivation for this study was concern about the fundamentals of
time-dependent density functional theory (TDDFT)\cite{MMNG12,U12}.  
Since the proof of a general
theorem\cite{RG84}, the number of applications of TDDFT in chemistry and physics has grown
phenomenally\cite{B12}.   
In its standard form, TDDFT translates the many-electron problem into a fictious
many-fermion problem {\em without} interaction between the particles, thereby
greatly reducing the computational cost, and allowing calculations with
several hundred atoms.
While all such applications rely on approximate functionals, their validity as an
alternative to solving the time-dependent Schr\"odinger equation relies on
several exact statements and the existence of exact functionals.

The most basic requirement for construction of a formally exact density functional
theory is a proof of uniqueness of the one-body potential that can give rise
to a given density.  The Runge-Gross theorem\cite{RG84} shows that, for a given
initial wavefunction and electron-electron repulsion, there is at most one $v(\br,t)$
that can produce a given $\n(\br,t)$ when solving TDSE.   Thus $v(\br,t)$ is a
functional of $\n(\br,t)$.  Applying the same logic to the fictitious KS system
yields the TD KS equations that can be applied to many-electron systems, once the
many-electron effects are approximated in the mysterious exchange-correlation
potential.  A linear response analysis\cite{GK85,C96,PGG96} yields an extremely
efficient scheme for calculating low-lying electronic excitations in molecules
and solids\cite{SDSG11}.

The proof of Runge-Gross was constructed only for one-body potentials that are
analytic in $t$, and can therefore be Taylor-expanded about $t=0$.  The proof
demonstrates that two distinct such potentials must give rise to densities
whose $n$-th derivative at $t=0$ differ for some finite $n$.  

The present and previous\cite{YMB12} work show that, in the case of a hydrogen atom
in a suddenly-switched electric field, the time-dependent density has non-analytic
contributions, so that the Taylor series does not converge.  Nonetheless, if
two densities differ in their $j$-th time-derivative, they must be different,
even if neither matches its Taylor expansion.  Thus the uniqueness proof of Runge-Gross
remains valid even for such problems.

This suggests these results apply to many-electron atoms, although they have only
been proven for one-electron cases.  If one considers the TD KS equations for, e.g., 
a He atom in a suddenly-switched field, in the region of the nucleus, the density,
which is a sum of occupied orbitals, will contain the same features (via the
occupied $1s$-orbital).  However, this argument presupposes the {\em existence}
of such a KS potential for this case.

Even in the simpler ground-state DFT, there are no general
conditions on densities known that guarantee that a density is in fact a ground-state
density for some electronic problem, although this is rarely a problem in practice,
even for strongly correlated systems\cite{SWWB12}.   

A second important theorem in TDDFT was van Leeuwen's constructive proof
of the TD KS potential.  Assuming both the density and the potential are Taylor-expandable,
a relatively simple procedure yields, power-by-power, a prescription 
for finding the potential\cite{L99}.   Clearly, this theorem does not apply to the
cases studied here.  Since all atoms, molecules and solids have cusps at their
nuclei (within the Born-Oppenheimer and point nuclei approximations), this theorem cannot be
applied as is to such cases.  Earlier work\cite{MTWB10} had already shown that
such cases could be constructed in 1d, but these could be regarded as pathological.
The motivation to develop the $s$-expansion described here was 
to convincingly show that such effects are generic, rather than unusual, once
the ground-state wavefunction contains spatial cusps.  In the last few years,
much work toward a proof of existence of the
KS potential without a Taylor-expansion has been performed\cite{RL11,RGPL12}.

\acknowledgements
ZY and KB thank Neepa Maitra for her help in the mathematical aspects, and Raphael Ribeiro for helpful discussions. ZY and KB acknowledge U.S. D.O.E funding (DE-FG02-08ER46496).


\begin{thebibliography}{10}

\bibitem{T07}
D.~J. Tannor,
\newblock {\em Introduction to quantum mechanics: a time-dependent
  perspective}
\newblock (University Science Books, 2007).

\bibitem{MMNG12}
\newblock {\em Fundamentals of time-dependent density functional theory},
\newblock edited by M.~A.~L. Marques, N.~T. Maitra, F.~M.~S. Nogueira, E.~K.~U. Gross, and
  A.~Rubio
\newblock (Lecture notes in physics. Springer, Berlin, 2012).

\bibitem{U12}
C.~A. Ullrich,
\newblock {\em Time-dependent density-functional theory: concepts and
  applications}
\newblock (Oxford University Press, Oxford, 2012).

\bibitem{EBF09}
P.~Elliott, F.~Furche, and K.~Burke,
\newblock In {\em Reviews in Computational Chemistry},
\newblock edited by K.~B. Lipkowitz and T.~R. Cundari
\newblock (Wiley, Hoboken, NJ, 2009).

\bibitem{ORR02}
G.~Onida, L.~Reining, and A.~Rubio,
\newblock Rev. Mod. Phys. {\bf 74}, 601 (2002).

\bibitem{BSDR07}
S.~Botti, A.~Schindlmayr, R.~Del~Sole, and L.~Reining,
\newblock Rep. Prog. Phys. {\bf 70}, 357 (2007).

\bibitem{RG84}
E.~Runge and E.~K.~U. Gross,
\newblock Phys. Rev. Lett. {\bf 52}, 997 (1984).

\bibitem{L99}
R.~van Leeuwen,
\newblock Phys. Rev. Lett. {\bf 82}, 3863 (1999).

\bibitem{YMB12}
Z.-H. Yang, N.~T. Maitra, and K.~Burke,
\newblock Phys. Rev. Lett. {\bf 108}, 063003 (2012).

\bibitem{RL11}
M.~Ruggenthaler and R.~van Leeuwen,
\newblock Eur. Phys. Lett. {\bf 95}, 13001 (2011).

\bibitem{RGPL12}
M.~Ruggenthaler, K.~J.~H. Giesbertz, M.~Penz, and R.~van Leeuwen,
\newblock Phys. Rev. A, {\bf 85}, 052504 (2012).

\bibitem{MTWB10}
N.~T. Maitra, T.~N. Todorov, C.~Woodward, and K.~Burke,
\newblock Phys. Rev. A, {\bf 81}, 042525 (2010).

\bibitem{AGD75}
A.~A. Abrikosov, L.~P. Gorkov, and I.~E. Dzyaloshinski,
\newblock {\em Methods of quantum field theory in statistical physics}
\newblock (Dover, New York, 1975).

\bibitem{FW03}
A.~L. Fetter and J.~D. Walecka,
\newblock {\em Quantum theory of many-particle systems}
\newblock (Dover, New York, 2003).

\bibitem{M95}
S.~Mukamel,
\newblock {\em Principles of nonlinear optical spectroscopy}
\newblock (Oxford, New York, 1995).

\bibitem{BF04}
H.~Bruus and K.~Flensberg,
\newblock {\em Many-body quantum theory in condensed matter physics}
\newblock (Oxford, New York, 2004).

\bibitem{HS72}
B.~R. Holstein and A.~R. Swift,
\newblock Amer.~J.~Phys. {\bf 40}, 829 (1972).

\bibitem{BO99}
C.~M. Bender and S.~A. Orszag,
\newblock {\em Advanced Mathematical Methods for Scientists and Engineers -
  Asymptotic Methods and Perturbation Theory}
\newblock (Springer, New York, 1999).

\bibitem{W10}
R.~B. White,
\newblock {\em Asymptotic analysis of differential equations}
\newblock (Imperial College Press, London, 2010).

\bibitem{AS72}
\newblock {\em Handbook of Mathematical Functions},
\newblock edited by M.~Abramowitz and I.~A. Stegun
\newblock (Dover, New York, 1972).

\bibitem{E06}
E.~N. Economou,
\newblock {\em Green's Functions in Quantum Physics}
\newblock (Springer, Berlin, 2006).

\bibitem{S95}
H.~J. Silverstone,
\newblock In {\em Modern elecronic structure theory},
\newblock edited by D.~R. Yarkony 
\newblock (World Scientific, Singapore, 1995).

\bibitem{F06}
H.~Friedrich,
\newblock {\em Theoretical Atomic Physics}, third edition
\newblock (Springer, Berlin, 2006).

\bibitem{MWB03}
N.~T. Maitra, A.~Wasserman, and K.~Burke,
\newblock In {\em Electron Correlations and Materials Properties 2},
\newblock edited by A.~Gonis, N.~Kioussis, and M.~Ciftan
\newblock (Kluwer Academic/Plenum Publishers, New York, 2003).

\bibitem{L01}
R.~van Leeuwen,
\newblock Int. J. Mod. Phys. B {\bf 15}, 1969 (2001).

\bibitem{BS57}
H.~A. Bethe and E.~E. Salpeter,
\newblock {\em Quantum Mechanics of One and Two-Electron Atoms}
\newblock (Springer, Berlin, 1957).

\bibitem{FC68}
U.~Fano and J.~W. Cooper,
\newblock Rev. Mod. Phys. {\bf 40}, 441 (1968).

\bibitem{YFB09}
Z.-H. Yang, M.~van Faassen, and K.~Burke,
\newblock J. Chem. Phys. {\bf 131}, 114308 (2009).

\bibitem{K57}
T.~Kato,
\newblock Commun. on Pure and Appl. Math. {\bf 10}, 151 (1957).

\bibitem{PTVF07}
W.~H. Press, S.~A. Teukolsky, W.~T. Vetterling, and B.~P. Flannery,
\newblock {\em Numerical Recipes}.
\newblock (Cambridge, third edition, 2007).

\bibitem{B12}
K.~Burke,
\newblock J. Chem. Phys. {\bf 136}, 150901 (2012).

\bibitem{GK85}
E.~K.~U. Gross and W.~Kohn,
\newblock Phys. Rev. Lett. {\bf 55}, 2850 (1985).

\bibitem{C96}
M.~E. Casida,
\newblock In {\em Recent developments and applications in density functional theory},
\newblock edited by J.~M. Seminario
\newblock (Elsevier, Amsterdam, 1996).

\bibitem{PGG96}
M.~Petersilka, U.~J. Gossmann, and E.~K.~U. Gross,
\newblock Phys. Rev. Lett. {\bf 76}, 1212 (1996).

\bibitem{SDSG11}
S.~Sharma, J.~K. Dewhurst, A.~Sanna, and E.~K.~U. Gross,
\newblock Phys. Rev. Lett. {\bf 107}, 186401 (2011).

\bibitem{SWWB12}
E.~M. Stoudenmire, L.~O. Wagner, S.~R. White, and K.~Burke,
\newblock Phys. Rev. Lett. {\bf 109}, 056402 (2012).

\end{thebibliography}
\end{document}